\documentclass[useAMS,usenatbib]{mnras}
\usepackage[english]{babel}

\usepackage{newtxtext,newtxmath}
\usepackage[T1]{fontenc}
\usepackage{ae,aecompl}

\usepackage{graphicx}	% Including figure files
\usepackage{amsmath}	% Advanced maths commands
\usepackage{amssymb}	% Extra maths symbols
\title[Metal and HI absorption lines at the end of the EoR]{Simulated metal and HI absorption lines at the conclusion of Reionization}

\author[L. A. Garc\'ia et al.]{
L. A. Garc\'ia$^{1,2}$\thanks{E-mail: lgarcia@swin.edu.au}, 
E. Tescari$^{2,3}$, E. V. Ryan-Weber$^{1,2}$ and J. S. B. Wyithe\,$^{2,3}$\\
\\
$^{1}$Centre for Astrophysics and Supercomputing, Swinburne University of Technology, Hawthorn, VIC 3122, Australia\\
$^{2}$ARC Centre of Excellence for All-Sky Astrophysics (CAASTRO)\\
$^{3}$School of Physics, The University of Melbourne, Parkville, VIC 3010, Australia
}

\date{Accepted 2017 June 01. Received 2017 May 02; in original form 2016 September 30}

\begin{document}
%\label{firstpage}
%\pagerange{\pageref{firstpage}--\pageref{lastpage}}

\maketitle

\begin{abstract} We present a theoretical study of intergalactic metal absorption lines imprinted in the spectra of distant quasars during and after the Epoch of Reionization (EoR). We use high resolution hydrodynamical simulations at high redshift ($4 <z<8$), assuming a uniform UV background Haardt--Madau 12, post-processing with {\small Cloudy} photoionization models and Voigt profile fitting to accurately calculate column densities of the ions \ion{C}{II}, \ion{C}{IV}, \ion{Si}{II}, \ion{Si}{IV} and \ion{O}{I} in the intergalactic medium (IGM). In addition, we generate mock observations of neutral Hydrogen (\ion{H}{I}) at $z<6$. Our simulations successfully reproduce the evolution of the cosmological mass density ($\Omega$) of \ion{C}{II} and \ion{C}{IV}, with $\Omega_{\ion{C}{II}}$ exceeding $\Omega_{\ion{C}{IV}}$ at $z >6$, consistent with the current picture of the tail of the EoR. The simulated \ion{C}{II} exhibits a bimodal distribution with large absorptions in and around galaxies, and some traces in the lower density IGM. We find some discrepancies between the observed and simulated column density relationships among different ionic species at $z=6$, probably due to uncertainties in the assumed UV background. Finally, our simulations are in good agreement with observations of the \ion{H}{I} column density distribution function at $z = 4$ and the \ion{H}{I} cosmological mass density $\Omega_{\ion{H}{I}}$ at $4 < z < 6$.  \end{abstract}


\section{Introduction}

Understanding the epoch of Reionization (EoR) is one of the current challenges of extragalactic as\-tro\-no\-my. It will complete the big picture of the thermal history of the Universe \citep[e.g.][]{mcquinn2015}. The EoR commenced when the first stars switched on, producing UV radiation that transformed the neutral Hydrogen (\ion{H}{I}) in the surrounding circumgalactic and  intergalactic media (CGM and IGM) into ionized hydrogen (\ion{H}{II}). Observations of Lyman--$\alpha$ photons (Ly$\alpha$ 1216 \AA) in absorption towards $z_{\text{em}} \gtrsim$ 6 quasars (QSOs) show fluctuations in flux consistent with Reionization concluding at $z \lesssim$ 6 \citep[e.g.][]{fan2006,becker2015}. Measuring Lyman series absorption or UV emissivity at wavelengths blueward of Ly$\alpha$ becomes almost impossible at redshifts greater than 5.5 due to the increasing density of matter and neutral Hydrogen fraction \citep{becker2015pasa}.\newline

Metal absorption lines are an alternative proxy to the Lyman series for probing the evolution of the IGM during the EoR and offer many advantages at high redshift. For example, they can be detected even when Hydrogen is completely saturated in the spectra of the background quasars, since some of these transitions occur redward of Ly$\alpha$ and are thus unaffected by Lyman series absorption and Gunn--Peterson troughs \citep{gunn1965}. Also, ionic transitions in the QSO spectra at high redshift provide important constraints on the model of the ionizing background that drove Reionization \citep[e.g.][]{furlanetto2009,becker2011,becker2015,finlator2016}. Low ionization transitions (\ion{O}{I}, \ion{C}{II}, \ion{Si}{II}, \ion{Mg}{II}, \ion{Fe}{II}) trace the location of neutral Hydrogen at high redshift. On the other hand, high ionization states (\ion{C}{IV}, \ion{Si}{IV}, \ion{O}{VI}) have a comparably larger ionization potential than H, therefore the energy required to produce these transitions is reached in regions where Hydrogen is highly ionized at early times.\newline The direct measurement of absorption features imprinted on the spectra of quasars at high redshift is the best method to infer the ionic ratios, but the observations depend on the wavelength bands where each transition can be found and the observational sensitivity is limited by the decreasing likelihood of an absorption line to be detected. In order to complement the observational techniques, hydrodynamical simulations are used to simulate the physical environment at high redshift where the absorption occurs. The numerical approach can produce a large number of sightlines, improving the statistical estimation of the column densities at redshifts that are not accessible due to current observational limits. Although the field is rapidly expanding, this approach is computationally expensive and the large dynamical range of the underlying physical phenomena makes a true self--consistent simulation impossible. Current simulations are still not able to simultaneously resolve the small scales (turbulence, shocks, fluctuations in the background density, etc.) and large scales (cosmological structures such as clusters and filaments) involved in the progression of the EoR. \newline An extensive observational effort has been pursued with \ion{C}{IV}, the triply ionized state of Carbon. \ion{C}{IV} offers many advantages that make its detection easier than other ionic species: large oscillator strength, wavelength redward of Ly$\alpha$ emission and a doublet transition. Observations of \ion{C}{IV} in the foreground of quasar spectra at high redshift by \citet{songaila2001, songaila2005}, \citet{pettini2003}, \citet{ryanweber2006, ryanweber2009}, \citet{simcoe2006}, \citet{simcoe2011}, \citet{dodorico2010, dodorico2013, dodorico2016} and \citet{boksenberg2015} have built the largest sample of metal absorbers as a function of redshift to date. The cosmological mass density $\Omega_{\ion{C}{IV}} = \frac{\rho_{\ion{C}{IV}}}{\rho_{\text{crit}}}$ (the density of \ion{C}{IV} with respect to the critical density today) drops at high redshifts during the progression of the EoR, due to the changing ionization state of the IGM and a decrease of its metallicity. \newline Numerical simulations by \citet{oppenheimer2006}, \citet{oppenheimer2009}, \citet{tescari2011}, \citet{cen2011}, \citet{pallottini2014}, \citet{finlator2015}, \citet{rahmati2016} and a comparison by \citet{keating2016} have tried to reproduce the evolution of $\Omega_{\ion{C}{IV}}$, taking into account different feedback prescriptions, photoionization modeling and variations in the UV ionizing background at high redshift. These theoretical efforts provide insight into the physical environments where \ion{C}{IV} absorptions occur and show that \ion{C}{IV} traces the distribution of the IGM at high temperature up to a few hundred kpc away from galaxies \citep[e.g.][]{oppenheimer2009}. \newline On the other hand, low ionization state ions, such as neutral oxygen \ion{O}{I}, should trace the distribution of \ion{H}{I} at high redshift. In fact, since the ionization potential of \ion{O}{I} differs from the neutral Hydrogen one by 0.02 eV, the two ions sit in tight charge-exchange. The detections of \ion{O}{I} toward high redshift QSOs in \citet{becker2006,becker2011} are consistent with large variations from one line of view to another, \textit{i.e.} with an inhomogeneous distribution of the ionizing sources as well as the absorbers. Numerical results from \citet{finlator2013} showed a tight correlation among \ion{H}{I} and \ion{O}{I} column densities and studies by \citet{keating2014} at N$_{\ion{H}{I}} >$10$^{17}$cm$^{-2}$ (where the gas is self-shielded) revealed an excellent agreement between \ion{O}{I} and \ion{H}{I} fractions, regardless of the photoionization model assumed, as well as an increasing incidence rate of \ion{O}{I} at higher redshift, consistent with the IGM being more neutral when approaching the EoR \citep{becker2011}.\newline Likewise, \ion{C}{II} has been observed by \citet{becker2006} and modelled by \citet{dodorico2013} at $z =$ 5.7 to be best fitted by low density gas, $\delta = \frac{\rho_{\text{gas}}}{\langle\rho\rangle}-1 = 10$, where ${\langle\rho\rangle}$ is the mean density at the considered redshift. Low ionization metals as \ion{C}{II} and \ion{Mg}{II} are routinely detected at low redshifts in regions close to the centre of galaxies or in damped Ly$\alpha$ systems \citep[DLAs, i.e. absorption systems with N$_{\ion{H}{I}} > 10^{20.3}$ cm$^{-2}$,][]{wolfe1986,wolfe2005}. These latter are optically thick structures that constitute the main reservoir of neutral Hydrogen after the EoR. DLAs have been studied both observationally \citep[e.g.][]{peroux2003, prochaska2005, omeara2007, prochaska2009, crighton2015} and theoretically/numerically \citep[e.g.][]{nagamine2004, pontzen2008, barnes2009, tescari2009, bird2014, rahmati2015, maio2015} in order to understand their physical properties and statistical distribution. However, many open questions still remain on the nature of these systems and their connection with the chemical enrichment of the Universe. \newline

The aim of this work is to i) compare high-$z$ observations of metal and \ion{H}{I} absorbers with synthetic spectra and ii) use simulations to gain insight into the underlying physical explanation for the observations. This paper is divided as follows: in section \ref{sims}, we introduce our numerical simulations. In section \ref{metho}, we explain the method used to include the UV background \citealt{haardt2012} (hereafter HM12 UVB), the photoionization modeling and the column density calculation procedure. In section \ref{igm}, we show the evolution with redshift of the physical properties that describe the IGM. Sections \ref{civ} and \ref{corr} analyze the ionic species in the simulations focussing on the cosmological mass density and the relationships among different ions. In the subsequent section \ref{hi}, we discuss the evolution of the cosmological mass density of neutral Hydrogen at $z<6$. Finally, in sections \ref{discussion} and \ref{conclusions} we summarize our results and the implications of the models. Throughout the paper we use the prefix c for comoving and p for physical distances.

\section{Cosmological simulations}
\label{sims}

The numerical simulations used in this work reproduce representative volumes of the Universe at redshift 4 $<z<$ 8 and were run with the smoothed particle hydrodynamics (SPH) code {\small P-GADGET3(XXL)} $-$ a customized version of {\small GADGET-3} \citep{springel2005}. The model is an extension at high redshift of the AustraliaN {\small GADGET-3} early Universe Simulations (\textsc{Angus}) project \citep{tescari2014, maio2015, katsianis2015, katsianis2016, katsianis2017}. Among other technical improvements, the suite of hydrodynamical simulations takes into account: a multiphase star formation criterion from \citet{springel2003}, self-consistent stellar evolution and chemical enrichment modeling \citep{tornatore2007}, supernova (SN) momentum-- and energy--driven galactic winds\footnote{Although active galactic nuclei (AGN) feedback is implemented in the code \citep{springel2005b,fabjan2010,planelles2013}, in this work we do not consider it. We stress that for the analysis presented in the paper the role of AGN feedback is expected to be negligible (see e.g. \citealt{keating2016} and references therein).}  \citep{springel2003,puckwein2013}, metal--line cooling \citep{wiersma2009} and low-temperature cooling by molecules/metals \citep{maio2007}. In addition, the code is supported by a parallel Friends-of-Friends (FoF) algorithm to identify collapsed structures and {\small SUBFIND} to classify substructures within FoF haloes.\newline Each simulation generates a cosmological box (including periodic boundary conditions) with the same initial number of gas and dark matter (DM) particles for a total of $2\times512^{3}$. Initial masses of the gas and DM particles are given in Table~\ref{table_sims}. Whenever the gas density is above a threshold $\rho_{\text{th}}$, there is a pro\-ba\-bi\-li\-ty that a gas particle will turn into a star\footnote{This probability is calibrated to reproduce the Kennicut-Schmidt law \citep{schmidt1959}.}. Stochastically, a new star--type particle is introduced in the simulation. Each star particle represents a simple stellar population with mass 0.1$M_{\sun} \leq m \leq$ 100 $M_{\sun}$. The stars with mass $m \leq$ 40 $M_{\sun}$ explode as SNe before turning into a black hole, while stars that have masses above this threshold collapse into a black hole without passing through the SN stage. \newline A flat $\Lambda$CDM model with cosmological pa\-ra\-me\-ters from the latest release of the \citet{planck2015} is assumed: $\Omega_{0m}=$ 0.307, $\Omega_{0b}=$ 0.049, $\Omega_{\Lambda}=$ 0.693, $n_s=$ 0.967, $H_0=$ 67.74 km s$^{-1}$Mpc$^{-1}$ (or $h =$ 0.6774) and $\sigma_8 =$ 0.816. The simulations were calibrated according to the parameters used in \citet{tescari2014} and \citet{katsianis2015}, and are compatible with observations of the cosmic star formation rate (SFR) density history and the galaxy stellar mass function at $z =$ 6 to 8.\newline The subgrid scheme takes into account the lifetimes of stars of different mass and follows the evolution of Hydrogen, Helium, molecules and metals (C, Ca, O, N, Ne, Mg, S, Si and Fe) released from SNIa, SNII and low and intermediate mass stars. It is possible to vary the initial mass function (IMF), the lifetime function and stellar yields. The algorithm is i\-dea\-lly suited to modeling IGM enrichment produced by galactic winds blown by ``starburst'' galaxies at high redshift. Radiative cooling and heating processes are included according to \citet{wiersma2009} and \citet{maio2007}.  The IMF considered is the Chabrier multi--sloped \citep{chabrier2003}, as described in equation \eqref{chabrier}, that produces a large number of intermediate-- and high--mass stars, expected to play an important role during Reionization.  \begin{equation}\label{chabrier} \zeta(m)= \begin{cases}
    \text{0.497} \times m^{-\text{0.2}} & \text{0.1} M_{\sun} \leq m < \text{0.3} M_{\sun} , \\
    \text{0.241} \times m^{-\text{0.8}} & \text{0.3} M_{\sun} \leq m < 1 M_{\sun} , \\
    \text{0.241} \times m^{-\text{1.3}} & m \geq 1 M_{\sun} .  \\
  \end{cases} \end{equation}
We adopted the following stellar yields:
\begin{itemize}
\item SNIa: \citet{thielemann03}. The mass range for the SNIa
  originating from binary systems is \mbox{0.8 $M_{\rm \odot}<m\le 8$
  $M_{\rm \odot}$}, with a binary fraction of 7\%.
\item SNII (massive stars): \citet{woosleyweaver95}. The mass range for SNII is
  $m>8$ $M_{\rm \odot}$.
\item Asymptotic giant branch (low and intermediate mass) stars: \citet{vandenhoek97}.
\end{itemize}

\begin{table*}
\centering
\caption{Summary of the simulations used in this work. Column 1: run name. Column 2: box size. Column 3: Plummer-equivalent comoving gravitational softening length. Columns 4 and 5: mass of gas and dark matter particles. All the simulations have the same initial number of gas and DM particles ($2\times512^3$). Column 6: feedback model. Column 7: inclusion of low-temperature metal and molecular cooling \citep{maio2007, maio2015}. The first run, Ch 18 512 MDW, is the fiducial model. The second one in the list, Ch 18 512 MDW mol, has exactly the same configuration as the reference run, but includes low-T metal and molecular cooling.}
\resizebox{0.95\textwidth}{!}{%
\label{table_sims}
\begin{tabular}{lcccccc} 
  \hline
  Simulation &  Box size & Comoving softening & $M_{\text{gas}}$ & $M_{\text{DM}}$ & Model for &  low-T metal \& \\
 &  (cMpc/$h$) & (ckpc/$h$) & ($\times$ 10$^{5}M_{\sun}$/$h$) & ($\times$ 10$^{6}M_{\sun}$/$h$) & SN--driven winds & molecular cooling\\ \hline
\textbf{Ch 18 512 MDW}  & \textbf{18} & \textbf{1.5} & \textbf{5.86} & \textbf{3.12} &\textbf{Momentum--driven} &    \\ 
Ch 18 512 MDW mol & 18 &1.5 &  5.86 & 3.12 & Momentum--driven &  \checkmark\\ 
Ch 18 512 EDW & 18 & 1.5 &  5.86 & 3.12 & Energy--driven &   \\ 
Ch 18 512 EDW mol& 18 & 1.5 & 5.86 & 3.12 &  Energy--driven &  \checkmark\\ 
Ch 12 512 MDW mol & 12 & 1.0 & 1.74 & 0.925  & Momentum--driven &  \checkmark\\ 
Ch 25 512 MDW mol & 25 & 2.0 & 15.73 & 8.48 & Momentum--driven &  \checkmark\\ 
\hline
\end{tabular}%
}
\end{table*}

%
%\begin{equation}
%\Phi (z) = \frac{\#_{\text{gal}}(\Delta M)}{V \cdot \Delta M},
%\end{equation}
%
%\begin{equation}
%V = (\text{box size(Mpc/h)})^3.
%\end{equation}
%\begin{equation}
%\text{log}(\Omega_{\text{CIV}} (z = 4)/\Omega_{\text{CIV}} (z = 8)) = 1.99 \sim 2
%\end{equation}
%
%\begin{equation}
%\text{log}(J_{\nu} (z = 4)/J_{\nu} (z = 8)) = 4.68 \:\:\:\:\:\:\:\:\: \text{at} \:\:\:\:\:\:\:\:\: \lambda_{\text{CIV}} =192 \AA
%\end{equation}
%
%\begin{equation}
%\text{log}(\Omega_{C} (z = 4)/\Omega_{C} (z = 8)) = 1.42
%\end{equation}

\subsection{Feedback mechanisms}

In order to regulate the star formation and chemically enrich the IGM, the simulations have been set up to account for kinetic supernova--driven winds. A\-ccor\-ding to the nature of the feedback and its effectiveness, galactic winds are classified as momentum-- or energy-- driven winds.

\subsubsection{Momentum--driven galactic winds (MDW)}

Momentum--driven galactic winds \citep{puckwein2013} are the reference feedback model used in this work. We assume that the mass-loss rate associated with the winds $\dot{M}_w$ is proportional to the star formation rate $\dot{M}_{\star}$ through $\eta$, the wind mass loading factor that accounts for the efficiency of the wind:
\begin{equation}\label{loading_fact}
\dot{M}_w = \eta \dot{M}_{\star}.
\end{equation}
In this model, the velocity of the  winds $v_w$ is regulated by the mass of the host halo and $R_{200}$, the radius where the density is 200 times larger than the critical density at redshift $z$:
\begin{equation}\label{wind-vel-mom}
v_w=2\sqrt{\frac{G M_{h}}{R_{200}}}=2 \times v_{\text{circ}},
\end{equation}
\noindent where $v_{\text{circ}}$ is the circular velocity and $R_{200}$ is defined as follows:
\begin{equation}\label{r200}
R_{200}=\sqrt[3]{\frac{3}{4\pi}\frac{M_h}{200\rho_c \Omega_{0m}}}(1+z)^{-1},
\end{equation}
\noindent with $\Omega_{0m}$ and $\rho_c$ the matter density and the critical density today. The conservation of momentum of the winds imposes that $\eta$ is proportional to the inverse of the wind--velocity $v_w$:
%To recover the dependence of $v_w$--$\eta$, the wind mass--loading factor is defined as:
\begin{equation}\label{eta}
\eta=2 \times \frac{\text{600 km s}^{-1}}{v_w}.
\end{equation}
\noindent The normalization factor was chosen to reproduce the observed cosmic SFR density and galaxy stellar mass function up to $z=8$. According to the model, weak winds ($v_w <$ 600 km s$^{-1}$) lead to a large efficiency and more material is expelled from supernovae with respect to strong winds ($v_w >$ 600 km s$^{-1}$) that reach further but load less material\footnote{Please note that in these simulations the normalization of the wind mass loading factor, $\eta_0 = 600$ km s$^{-1}$, is slightly larger than in previous \textsc{Angus} runs, such as those presented in \citet{tescari2014} and \citet{katsianis2015}, where $\eta_0 = 450$ km s$^{-1}$. This is due to the fact that these simulations do not include AGN feedback (and in particular the early AGN feedback in low-mass galaxies model), and therefore a recalibration of the wind model was necessary for properly matching the cosmic star formation rate density history and galaxy stellar mass function up to $z=8$.}. Stochastically, some gas particles are selected to be part of the wind, and subsequently, decouple from the hy\-dro\-dy\-na\-mi\-cal scheme for a given amount of time $t_{\text{dec}}$ \citep{tescari2014}.

\subsubsection{Energy--driven galactic winds (EDW)}

To test the impact of the adopted feedback model on our results, we also use energy--driven winds \citep{springel2003}. The only remarkable difference with momentum--driven winds is the scaling form of $\eta$. As for MDW, the speed of the winds is regulated by $v_w= 2 \times v_{\text{circ}}$, but in this case due to conservation of the energy of the winds, the wind mass--loading factor $\eta$ scales as the square of the inverse of $v_w$,
\begin{equation}\label{eta}
\eta=2 \times \left(\frac{\text{600 km s}^{-1}}{v_w}\right)^2,
\end{equation}
\noindent making EDW more aggressive than MDW, especially in low-mass galaxies. Thermal feedback produced by SNIa and SNII is also considered, in addition to the kinetic feedback just described.

\section{Methodology} \label{metho}

The set of simulations was run to $z =$ 4, with initial conditions at $z =$ 125, using the Raijin supercluster from the National Computational Infrastructure (NCI) facility\footnote{http://nci.org.au}. These simulations describe the physical conditions of the gas and the evolution of H, He, C, Ca, O, N, Ne, Mg, S, Si and Fe released from SNIa and SNII. We post-process the simulations introducing a uniform UV ionizing background: a field radiation due to the CMB and the \citet{haardt2012} ultraviolet/X-ray background from quasars and galaxies with saw-tooth a\-tte\-nua\-tion that evolves with redshift (hereafter HM12).\newline
Assuming this UVB, we compute the ionization states of each element with {\small Cloudy} photoionization code v8.1 \citep{ferland2013} for optically thin gas in ionization e\-qui\-li\-brium, focusing in particular on the following ions: \ion{H}{I},  \ion{C}{II}, \ion{C}{IV}, \ion{Si}{II}, \ion{Si}{IV} and \ion{O}{I}. We consider only the transition with the highest oscillator strength and rest-frame wavelength $\lambda_{\text{rest}} \geq$ 1216 \AA, which can be observed at high redshift. The assumed metallicity in the {\small Cloudy} tables is solar. \newline
In addition, an effective prescription for \ion{H}{I} self-shielding was introduced to accurately describe the regions at the centre of the galaxies, where N$_{\ion{H}{I}}$ is significantly higher than in the IGM due to the shielded bubbles that contain pristine Hydrogen. We adopt the parametric function of \citet{rahmati2013a}:
\begin{equation}\label{gamma_phot}
\frac{\Gamma_{\text{phot}}}{\Gamma_{\text{UVB}}}=(1-f)\left[1+\left( \frac{n_{\text{H}}}{n_0}\right)^{\beta} \right]^{\alpha_1}+f \left[1+\frac{n_{\text{H}}}{n_0}\right]^{\alpha_2} ,
\end{equation}
\noindent which best--fitting parameter values reproducing the radiative transfer results at $1<z<5$ are: $\alpha_1=-2.28\pm 0.31$, $\alpha_2=-0.84\pm0.11$, $\beta=1.64\pm 0.19$, $n_0=(1.003\pm 0.005)$ $n_{\text{H,SSh}}$ (where $n_{\text{H,SSh}}$ is the self-shielding density threshold) and $f= 0.02\pm 0.0089$. The equilibrium neutral Hydrogen density is obtained using a recombination rate given by \citet{hui1997} and the photoionization rate as a function of the temperature of \citet{theuns1998}. \newline
In order to mimic real observations and avoid the introduction of bias in the data, we generate random lines of sight inside the simulated box along the three perpendicular directions and extract the relevant physical information taking into account positions, velocities, densities and tem\-pe\-ra\-tu\-res of the SPH particles inside each line of sight. Subsequently, we compute a synthetic spectra in density and optical depth/flux as a function of the velocity width (1024 pixels) for each ion in the simulation, according to the procedure introduced by \citet{theuns1998}. For a bin $j$ at a position $x(j)$, the density and density weighted temperature and velocity are calculated from:
\begin{equation}\label{rhox}
\rho_X(j) =  a^3 \sum_i X(i) W_{ij},
\end{equation}
\begin{equation}\label{rhotx}
(\rho T)_X(j) =  a^3 \sum_i X(i) T(i) W_{ij},
\end{equation}
\begin{equation}\label{rhovx}
(\rho v)_X(j) =  a^3 \sum_i X(i) (a\dot{x}(i)+\dot{a}\left[x(i)-x(j)\right]) W_{ij}, 
\end{equation}
\noindent where $a$ is the scale factor, $X(i)$ the abundance of the species $X$ of the SPH particle $i$ and $W_{ij}=mW(q_{ij})/h_{ij}^3$ the nor\-ma\-li\-zed SPH kernel. $W$ is the SPH kernel, $m$ the particle mass and:
\begin{equation}\label{qij}
q_{ij} = \frac{a|x(i)-x(j)|}{h_{ij}},
\end{equation}
\begin{equation}\label{hij}
h_{ij} =\frac{1}{2}\left[h(i)+h(j)\right],
\end{equation}
\noindent with $h$ the physical softening scale.\newline
With this information, it is possible to calculate the number density of the ion transition considered $n_{\text{ion}}$. The synthetic flux for any transition at the redshift-space coordinate $u$ is given by $F(u)=\text{exp}\left[-\tau(u)\right]$, with $\tau(u)$:
\begin{equation}\label{tau}
\tau(u) = \frac{\sigma_{0,I} c}{H(z)}\int_{-\infty}^{\infty} n_{I}(x)V\left[u-x-v_{\text{pec,||}}^{\text{IGM}}(x), b(x)\right] dx,
\end{equation}
\noindent where $\sigma_{0,I}$ is the cross-section of the ion transition, $H(z)$ the Hubble parameter at $z$, $x$ the space coordinate in \mbox{km s$^{-1}$}, $b$ the velocity dispersion (in units of $c$) and $V$ the Voigt profile. The spectra can be converted from the velocity space $v$ to the observed wavelength using $\lambda = \lambda_0 (1+z)(1+v/c)$.\newline In numerical works, it is a common practice to normalize \ion{H}{I} fluxes averaged over a large number of random lines of sight to the observed mean normalized flux of the Ly$\alpha$ forest at a given redshift, $\langle F(z)\rangle = \exp(-\tau_{\text{eff}})$, through a constant rescaling factor $A_{\ion{H}{I}}$. For consistency, also the fluxes of metal ions are rescaled by the same constant factor. However, at $z>5$ reliable measurements of $\tau_{\text{eff}}$, and therefore $A_{\ion{H}{I}}$, are not possible, due to the thickening of the Ly$\alpha$ forest. To overcome this problem, we adopted a reversed approach. First, we calculated the \ion{C}{IV} column density distribution function at $z = $ 4.8 and 5.6 (see section \ref{c4cddf}) using three different simulations (Ch 18 512 MDW, Ch 18 512 MDW mol and Ch 18 512 EDW) and a range of scaling factors for the \ion{C}{IV} optical depths, $\tau_{\ion{C}{IV}}$. Then, we selected the scaling factor, $A_{\ion{C}{IV}}$, which provided the best chi-by-eye agreement between the simulated and observed distribution functions. Finally, we rescaled all the ionic fluxes (including \ion{H}{I}) in all the simulations and at all the redshifts considered (i.e. also at $z\le5$, for consistency) by $A_{\ion{C}{IV}}$. The best fit value is $A_{\ion{C}{IV}}=0.85$, very close to unity.\newline The individual spectra are convolved with Gaussian noise profiles with full width at half maximum $\text{FWHM} =$ 7 \mbox{km s$^{-1}$} to produce final synthetic spectra with a signal-to-noise ratio, $S/N=$ 50, comparable with observations obtained using the UVES spectrograph mounted at the Unit 2 of the Very Large Telescope. Finally, to obtain a fair com\-pa\-ri\-son with real data, we fit all the individual absorption features in the spectra through Voigt profile components in a range of 100 \mbox{km s$^{-1}$} among contiguous systems with the code {\small Vpfit} v.10.2 \citep{vpfit}. \newline The column density N, equivalent width $EW$ and Doppler parameter $b$ from each spectrum is used to produce a sample for each ion. The estimated errors of the column density $\Delta$N, Doppler parameter $\Delta$b and redshift $\Delta$z of the absorption features are also calculated with {\small Vpfit}. The individual fits are selected if the conditions N $> \Delta$N and \mbox{b $> \Delta$b} are fulfilled simultaneously. Otherwise, the component is rejected.\newline

As a final remark, we note that radiative transfer effects are not included in our simulations. The models use the evolving HM12 uniform UVB to quantify the ionization state of the CGM/IGM at a given redshift. The aim of this work is not to follow the progression of Reionization, nor the evolution of the \ion{H}{II} bubbles or their topology. The implicit assumption is that our boxes (that are small compared to the size of the \ion{H}{II} bubbles at the redshifts of interest) represent a region of the Universe already reionized at a level given by the HM12 UVB. At $6 < z < 8$, chemical enrichment occurs mostly inside and in close proximity of galaxies (interstellar medium, CGM and high density IGM) where, assuming an inside-out progression of Reionization, the gas in which metals lie should be ionized. Although proper RT calculations would be more accurate, they are extremely expensive from the computational point of view. Using a uniform (evolving) UVB is a common approach in the literature \citep[see e.g.][]{oppenheimer2009, keating2016}. Moreover, \citet{finlator2015} studied the evolution of Carbon absorption from $z = 10$ to $z = 5$ with cosmological hydrodynamic simulations that include a self-consistent multifrequency, inhomogeneous UVB. They found that the difference between their more realistic UVB and the uniform HM12 is within $\sim2-4$ times, which according to them is fair agreement given the uncertainties.\newline We stress that in this work we do not present results on the neutral Hydrogen fraction at $z > 6$. In this case, a uniform background slowly ionizing the IGM is a completely wrong picture. For this reason, we only study the \ion{H}{I} fraction at $z < 6$. As already mentioned before, to accurately describe the regions at the centre of galaxies we also include an effective prescription for \ion{H}{I} self-shielding \citep{rahmati2013a}.

\section{Evolution of the IGM during Reionization}
\label{igm}

\begin{figure*}
\includegraphics[scale=0.25]{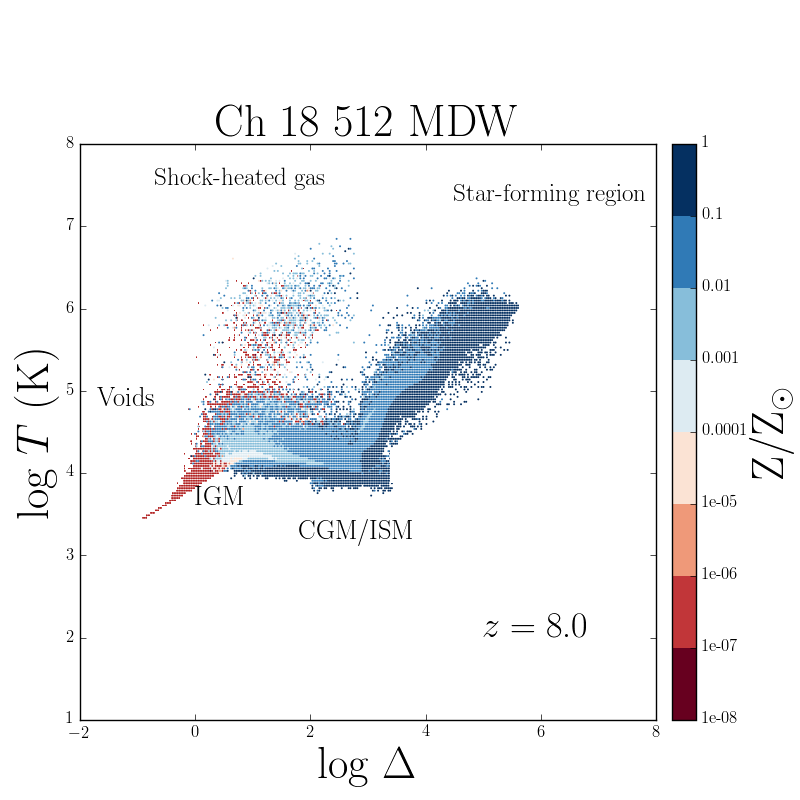}
\includegraphics[scale=0.3]{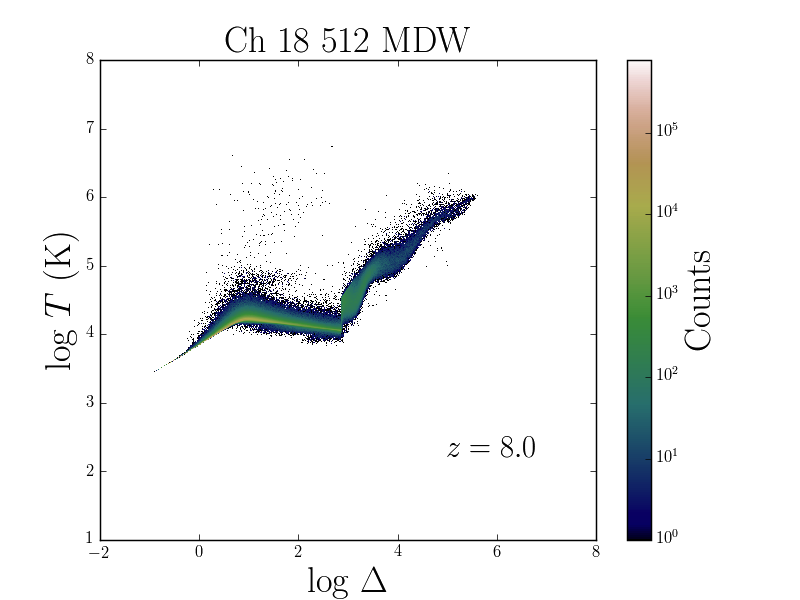}
\includegraphics[scale=0.28]{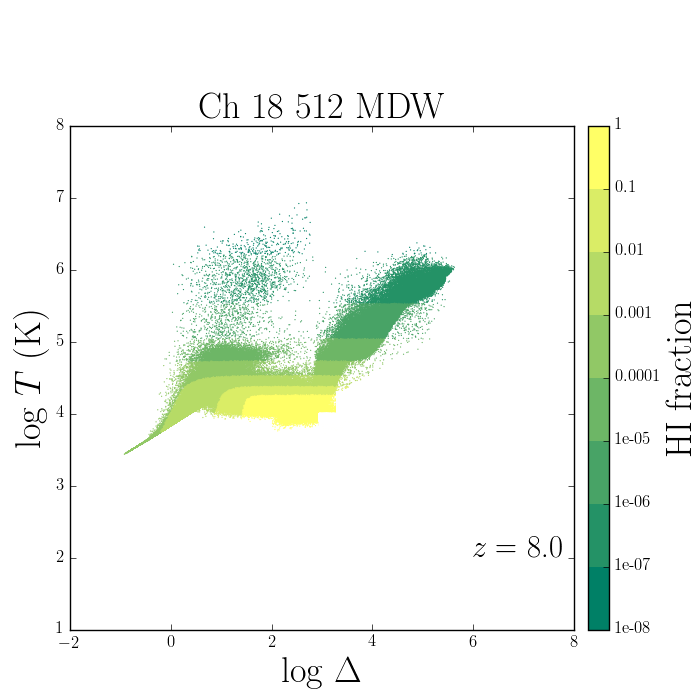}
%\vspace{0.0cm}
\hspace{0.0cm} \includegraphics[scale=0.25]{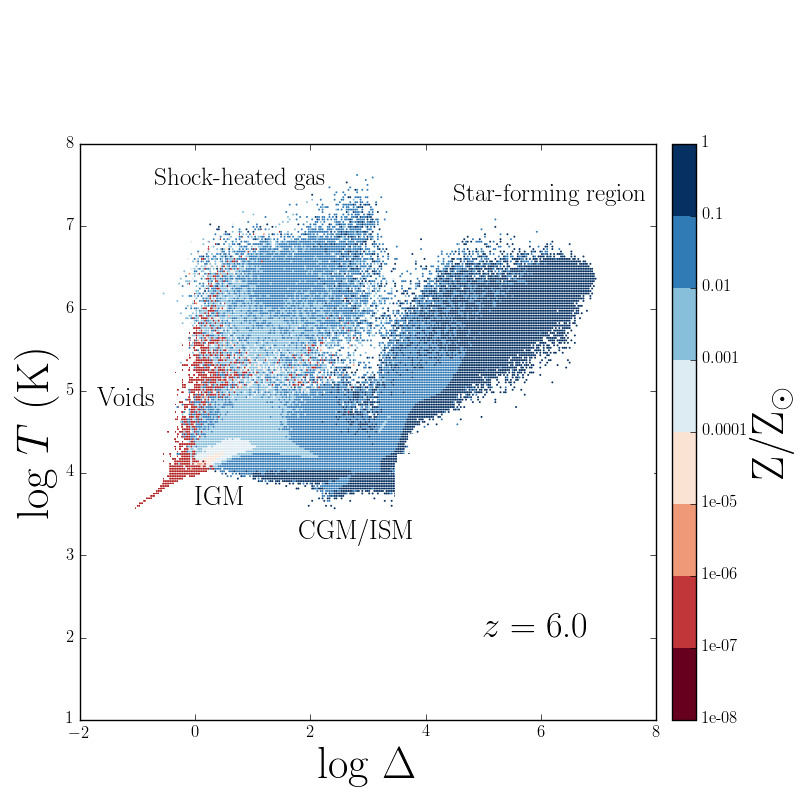} \includegraphics[scale=0.3]{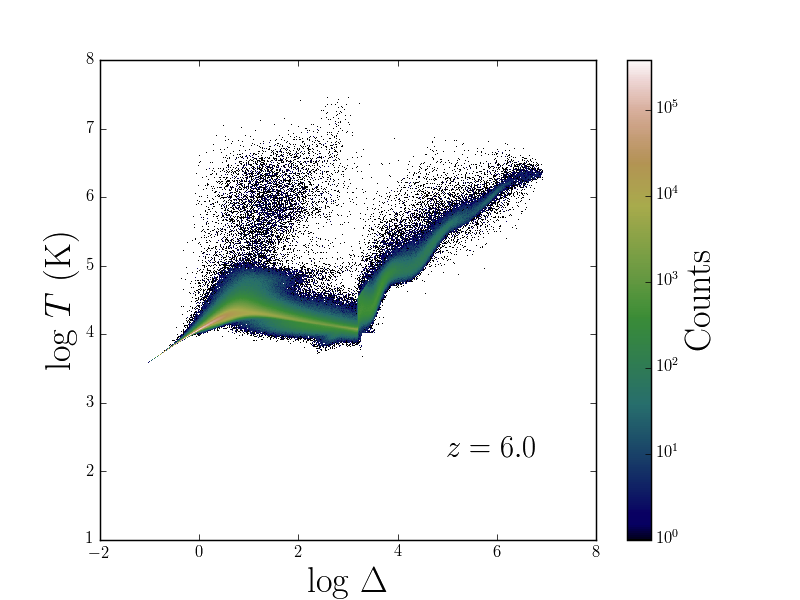} \includegraphics[scale=0.28]{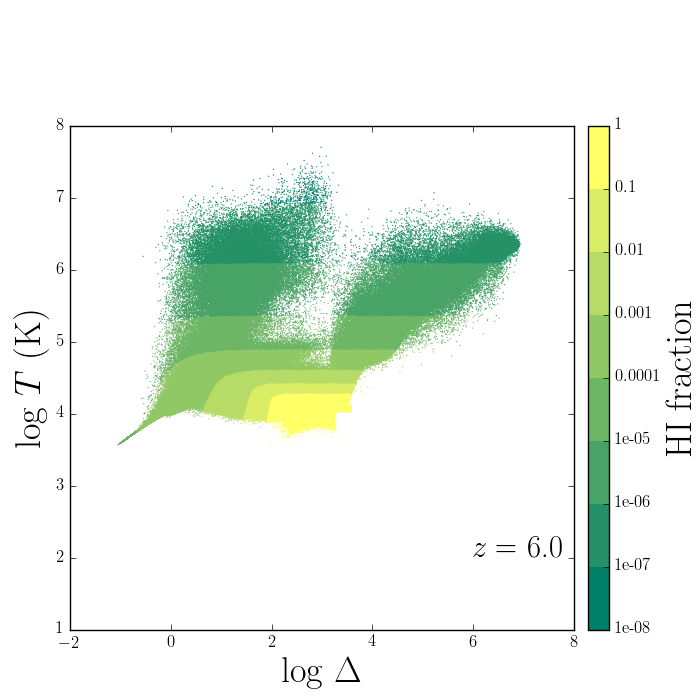} \vspace{0.0cm} \hspace{0.0cm} \includegraphics[scale=0.25]{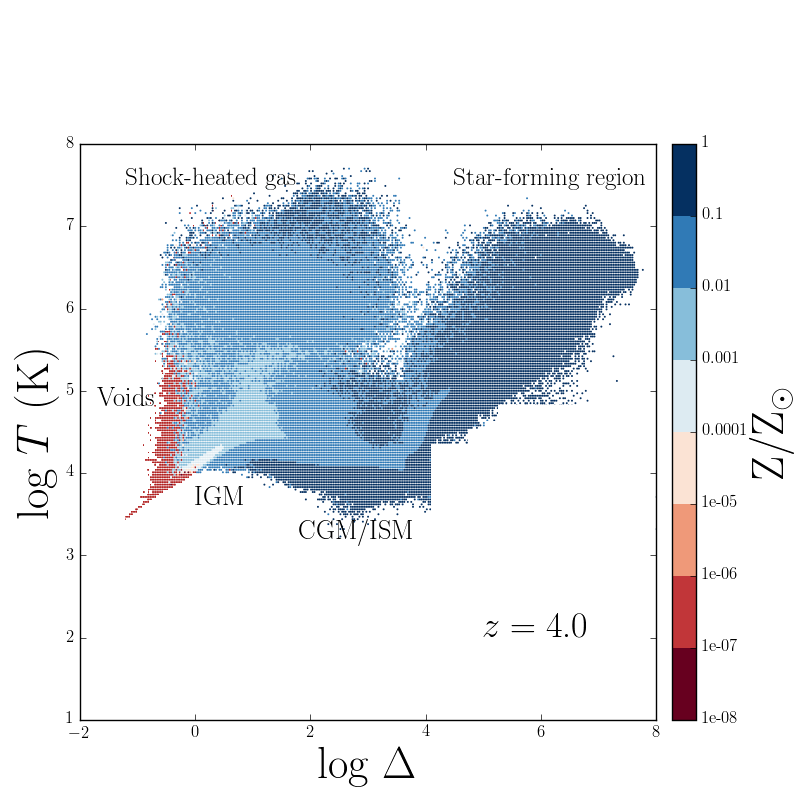} \includegraphics[scale=0.3]{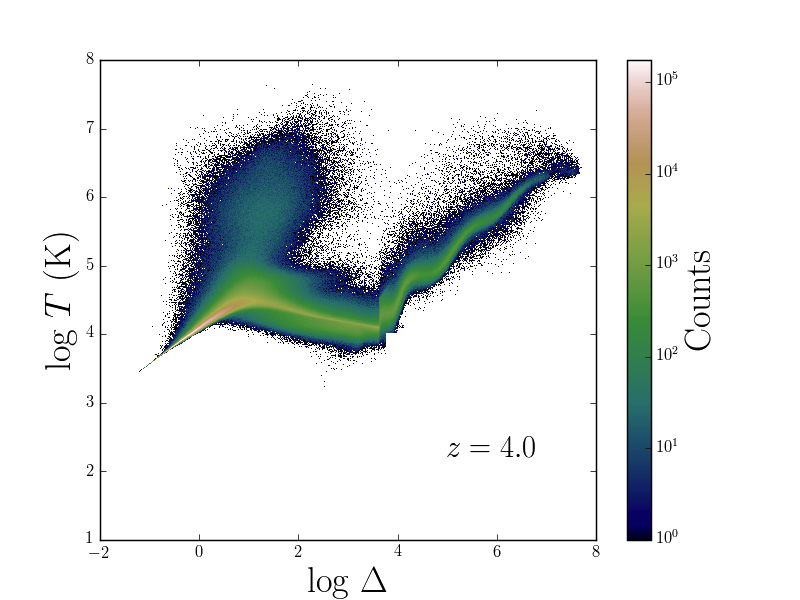} \includegraphics[scale=0.28]{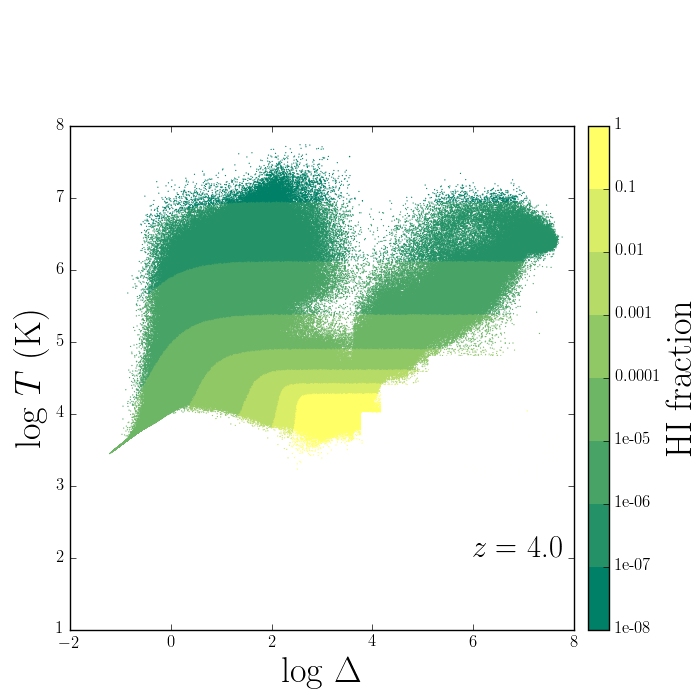} \vspace{0.0cm} \caption{Density--temperature diagram for gas particles at redshifts $z =$ 8, 6 and 4 for the fiducial model Ch 18 512 MDW. The left panels corresponds to the distribution of the total metallicity (in solar units). The dark red region represents zero metallicity. As a reference, the different regions are tagged (voids, IGM, CGM/ISM, shock--heated gas and star--forming region). In the middle panels, the colour maps represent the number of particles with overdensity and temperature in a given ($\log \Delta_i$, $\log T_i$) bin. On the right hand side, we show the distribution of \ion{H}{I} fraction. The top right panel was included for illustrative purposes and does not represent a realistic description of the neutral Hydrogen content of the Universe at $z=8$ (see the discussion at the end of section \ref{metho}).}
\label{fig:scatter_plot}
\end{figure*}
Figure~\ref{fig:scatter_plot} shows the distribution of the gas particles in the fiducial simulation Ch 18 512 MDW in diagrams of overdensity--temperature at redshifts $z=$ 8, 6 and 4, comparing total metallicity, number of particles and \ion{H}{I} fraction (left, middle and right panels, respectively). These redshifts are chosen to study the IGM at i) $z =$ 8 when Reionization is progressing; ii) $z =$ 6, when the overlap of the \ion{H}{II} bubbles is nearly complete at the end of the EoR; and iii) $z =$ 4 when the Reionization of Hydrogen has fully concluded and the only remaining reservoirs of neutral Hydrogen are self-shielded regions of \ion{H}{I} and DLA systems.\newline
The diagrams are split in well--defined regions whose sizes evolve with time in terms of the overdensity, $\Delta = \delta +1 = \frac{\rho_{\text{gas}}}{\langle\rho\rangle}$, and the temperature $T$:
\begin{itemize}
\item voids: gas particles at very low densities ($\Delta < 1$);
\item IGM: the gas in this regime follows an adiabatic relation $T=T_o \Delta^{\gamma-1}$ and is in ionization equilibrium: cooling is counterbalanced by photoheating. The conditions of the particles in the IGM are $T \leq$ 10$^4$ K and $1 \leqslant \Delta < 10$;
\item Circumgalactic/interstellar media (CGM/ISM) represent the transition between the IGM and the inside of galaxies, with typical densities of $\Delta \sim$ 10$^{1-3}$. This environment is heated by photoionization fronts coming from galaxies. The sharp feature at $\Delta \sim$ 10$^{3}$ represents the density threshold where gas particles turn stochastically into star--particles in the simulation;
\item star--forming region: particles in this regime are above $\Delta \sim$ 10$^{3}$, have temperatures higher than 10$^{4}$ K and follow an effective equation of state imposed by the subgrid star formation model \citep{springel2003};
\item shock--heated gas: a growing region of the IGM that is heated by feedback processes at late times ($T >$ 10$^{5}$ K and low densities).
\end{itemize}
At $z =$ 6 -- 8, most of the gas in the diagrams is located in the so--called diffuse phase $-$ low $T$ ($\leq$ 10$^4$ K) and low $\Delta$ ($\leqslant$ 10$^2$) $-$ the first stars are being formed and the main processes that raise the temperature of the IGM (and voids) are taking place.\newline The leftmost panels show the metallicity with respect to solar. The enriched sections of the phase diagram are mainly star-forming regions and some gas at very high temperature, expelled from galaxies through supernova--driven winds. On the other hand, voids exhibit zero metallicity, because the galactic outflows do not reach regions that far from the centre of the galaxies. The CGM and ISM are chemically enriched at high redshift. Instead, the IGM is very metal-poor at $z =$ 8. At late stages (bottom panels), feedback prescriptions spread out material from the high-density gas to regions originally empty at higher redshifts. Chemical enrichment contributes progressively more to the cooling processes.\newline The right panels show the distribution of Hydrogen in the reference simulation. Yellow indicates Hydrogen completely neutral ($X_{\text{HI}} \geqslant $ 10$^{-1}$), dark green particles containing \ion{H}{II} ($X_{\text{HI}} <$ 10$^{-7}$) and the intermediate colors are used to illustrate the transition between the two regimes. It is clear that most of the reservoirs of \ion{H}{I} are located at low temperature and intermediate-- to high--density regions, that are presumably where DLAs or isolated self-shielded regions are. On the other hand, the shock--heated gas with tem\-pe\-ra\-tu\-res $T >$ 10$^5$ K contains a large amount of ionized gas. The fraction of this gas that has been chemically enriched is expected to be found in high ionization states, such as \ion{C}{IV} or \ion{Si}{IV}, consistent with \citet{cen2011}. \newline Please note that, for the reasons discussed at the end of section \ref{metho}, the top right panel should not be regarded as an accurate representation of the neutral Hydrogen content in the Universe at $z=8$, since a uniform UVB does not properly describe the progression of Reionization. We included this panel just for illustrative purposes.

\section{Evolution of Carbon}
\label{civ}

In order to study the evolution of the IGM at high redshift, we use synthetic spectra and analyze the evolution of metal absorption lines.

\subsection{CIV column density distribution function}
\label{c4cddf}

\ion{C}{IV}, the triply ionized state of Carbon, is the transition most detected and studied in the foreground of quasar spectra at high redshift. We devote this section to its cosmological evolution and compare our theoretical predictions with the available observational archive.\newline

\noindent The column density distribution function (CDDF) is an observable that takes into account the statistical distribution of absorption systems with respect to their column densities. Once the absorptions are detected and the redshift path is confirmed to be complete, the CDDF can be used to build the cosmological mass density of a particular ion.
By construction, the CDDF, or $f(\text{N},X)$, quantifies the number of absorption systems $n_{\text{sys}}$ in the column density interval $(\text{N}, \text{N}+\Delta \text{N})$ in an absorption path $\Delta X$:
\begin{equation}\label{CDDF_ion}
f(\text{N},X)=\frac{ n_{\text{sys}}(\text{N}, \text{N}+\Delta \text{N})}{n_{\text{lov}} \Delta X},
\end{equation}
\noindent where $n_{\text{lov}}$ is the number of lines of view (lov) considered. The absorption path relates the Hubble parameter at a given redshift $z$ with the correspondent redshift path $\Delta z$ as follows:
\begin{equation}\label{dX}
\Delta X= \frac{H_0}{H(z)}(1+z)^2 \Delta z.
\end{equation}
\noindent The equivalent redshift path of our cosmological boxes is defined as:
\begin{equation}\label{dz}
\Delta z =  (1+z)\frac{\Delta v}{c},
\end{equation}
\noindent where $\Delta v$ is the box size in km s$^{-1}$ at a given redshift $z$.

\begin{figure}
	\includegraphics[width=\columnwidth]{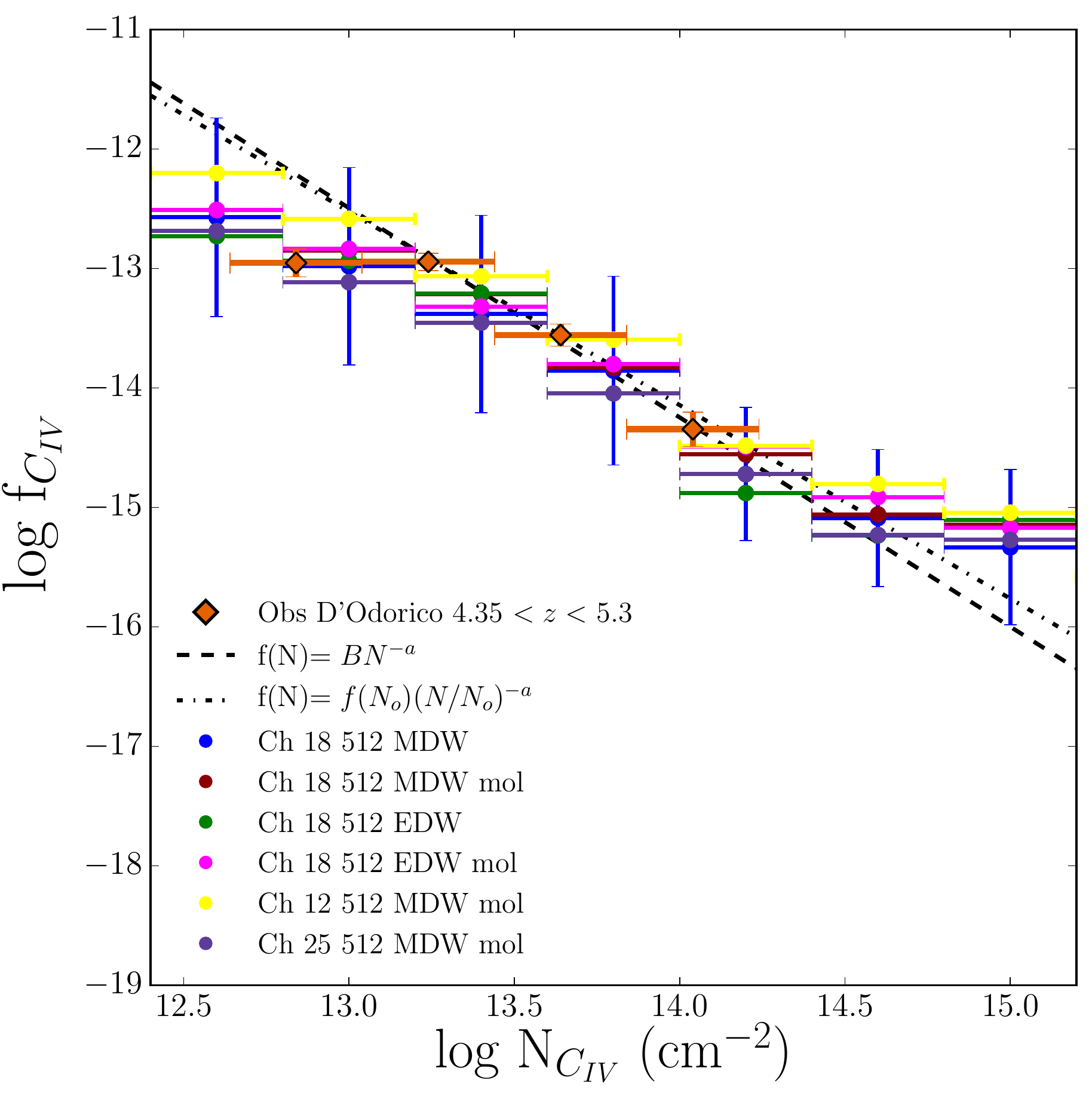}
        \caption{\ion{C}{IV} column density distribution function at $z =$ 4.8 and comparison with observational data by \citet{dodorico2013} in orange diamonds. The black dashed line represents the fitting function $f($N$) = B$N$^{-\alpha}$ with $B = $ 10.29 $\pm$ 1.72 and $\alpha =$ 1.75 $\pm$ 0.13 and the dotted--dashed line $f($N$) = f($N$_0)($N$/$N$_0)^{-\alpha}$ with $f($N$_0) = $ 13.56 and $\alpha =$ 1.62 $\pm$ 0.2, from the same observational work. The blue error bars are the Poissonian errors for the reference run and are a good representation of the errors in the other models. Hereafter, these colors are used to represent the simulations.}
    \label{fig:CIVCDDF4.8}
\end{figure}

\begin{figure}
	\includegraphics[width=\columnwidth]{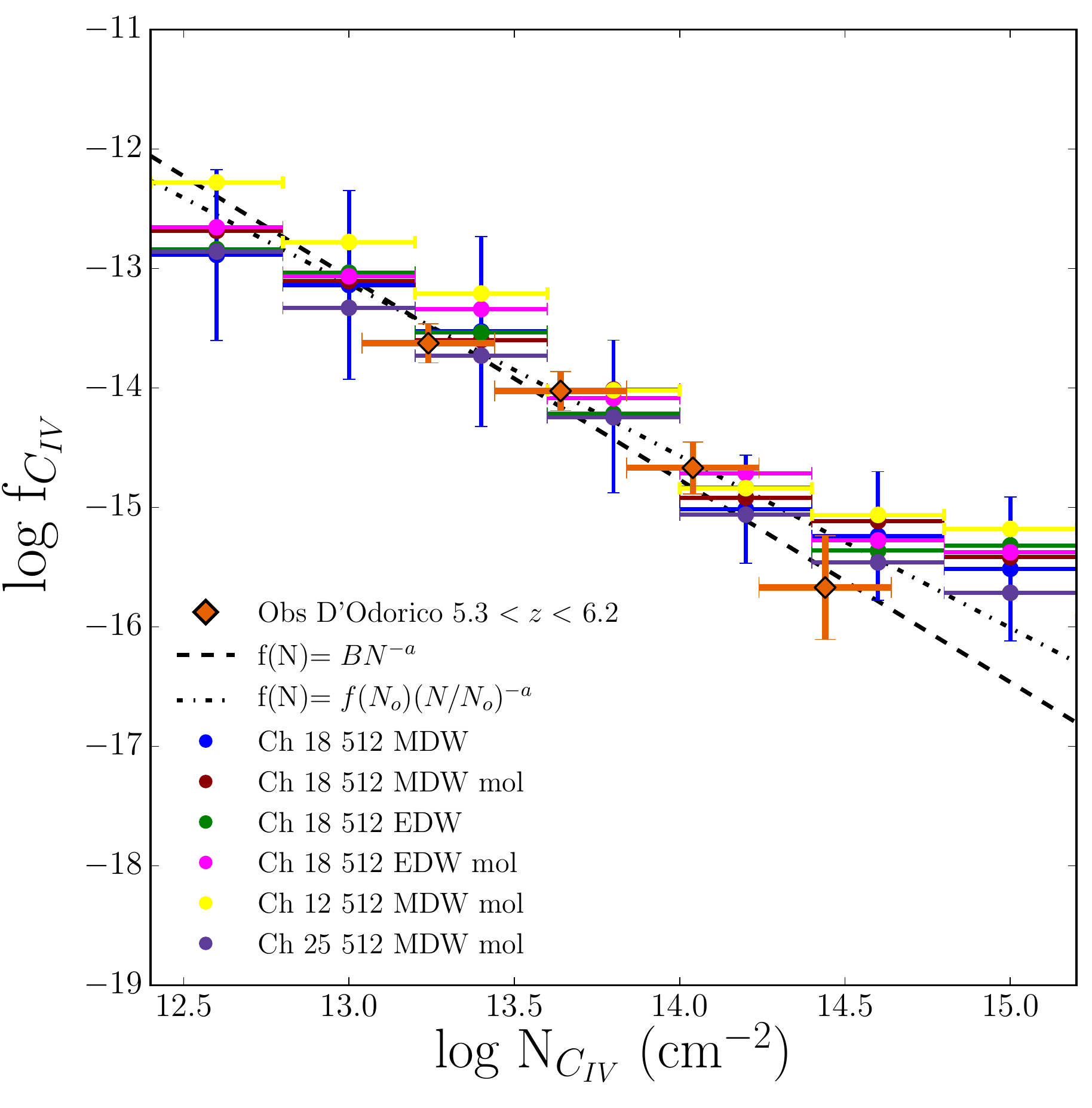}
        \caption{\ion{C}{IV} column density distribution function at $z =$ 5.6 and comparison with observational data by \citet{dodorico2013} in orange diamonds. The black dashed line represents the fitting function $f($N$) = B$N$^{-\alpha}$ with $B = $ 8.96 $\pm$ 3.31 and $\alpha =$ 1.69 $\pm$ 0.24 and the dotted--dashed line $f($N$) = f($N$_0)($N$/$N$_0)^{-\alpha}$ with $f($N$_0) = $ 14.02 and $\alpha =$ 1.44 $\pm$ 0.3, from the same observational work. The blue error bars are the Poissonian errors for the reference run and are a good representation of the errors in the other models. At large column densities, simulations predict values of the \ion{C}{IV}--CDDF slightly higher than observations. However, the error bars on all of the theoretical models overlap with the observational data.}
    \label{fig:CIVCDDF5.6}
\end{figure}

\noindent We impose a cut--off in column density, $\log$ N$_{\text{th}}$ (cm$^{-2}$) $=$ 12.5, to the synthetic data to mimic the sensitivity of available observations and to avoid including poorly sampled features. Although a large number of systems below this column density threshold is present ($\sim80$\%), these have not been taken into account in the statistics. Moreover, we limit the calculation of the \ion{C}{IV}--CDDF to \mbox{N$_{\text{\ion{C}{IV}}}$ (cm$^{-2}$) $<$ 10$^{15.2}$}, consider bins of 0.4 dex and introduce Poissonian errors for the theoretical sample as $\sqrt{n_{\text{sys}}}$ to fairly compare with the observations of \citet{dodorico2013}.\newline Figure~\ref{fig:CIVCDDF4.8} shows the predicted column density distribution function of \ion{C}{IV} at $z =$ 4.8 and compares with observations by \citet{dodorico2013} and two fitting functions proposed by the authors: $f($N$) = B$N$^{-\alpha}$ with $B = $ 10.29 $\pm$ 1.72 and $\alpha =$1.75 $\pm$ 0.13 and $f($N$) = f($N$_0)($N$/$N$_0)^{-\alpha}$ with $f($N$_0) = $ 13.56 and $\alpha =$ 1.62 $\pm$ 0.2. Although we used this statistics as a rough guide to calibrate the \ion{C}{IV} optical depths in three of our simulations (see the final part of section \ref{metho}), there is good agreement among {\it all} the simulations and the observational data at this redshift, and small deviations from the observations are within the error bars. The error bars in \ion{C}{IV}--CDDF shown correspond to the reference run and are representative for all the simulations.\newline The same observable is computed at $z =$ 5.6 in Figure~\ref{fig:CIVCDDF5.6}. The values from the numerical simulations are shown in points and they are compared to systems detected by \citet{dodorico2013}. The observational fitting functions parameters for $f($N$) = B$N$^{-\alpha}$ are $B = $ 8.96 $\pm$ 3.31, $\alpha =$ 1.69 $\pm$ 0.24 and for $f($N$) = f($N$_0)($N$/$N$_0)^{-\alpha}$ are $f($N$_0) = $ 14.02 and $\alpha =$ 1.44 $\pm$ 0.3. There is good agreement between the observational and theoretical points except at large column densities, where simulations deviate to values higher than observations. We point out that the function fits at $z =$ 4.8 and 5.6 extend to high column densities, even though observations are only available up to N$_{\text{\ion{C}{IV}}} =$ 10$^{14.4}$ cm$^{-2}$. Therefore, the outcome of the simulations gives additional information in a range of N$_{\ion{C}{IV}}$ where the \ion{C}{IV}--CDDF has not been measured.\newline We stress that, at these redshifts, the \ion{C}{IV}--CDDF cannot be used to disentangle different physical prescriptions in the simulations, since all the runs are in agreement within the error bars.\newline Finally, it is important to note that the run Ch 12 512 MDW mol predicts the highest \ion{C}{IV}--CDDF at low column density both at $z =$ 4.8 and 5.6. The higher resolution of this run with respect to all the others leads to a more complete and precise estimation of the column density distribution function at low N$_{\text{\ion{C}{IV}}}$. \newline

In similar works by \citet{finlator2015,finlator2016} and \citet{keating2016}, strong absorbers are highly disfavored for different simulated UV backgrounds and feedback prescriptions. In \citet{finlator2015,finlator2016}, the feedback me\-cha\-nisms are quite efficient at enriching the IGM with metals. However, their simulated boxes are relatively small. Therefore, galaxies with large halo masses are suppressed. If these structures host the high column density systems, such simulations will also lack strong absorptions. On the other hand, \citet{keating2016} explore various feedback prescriptions. Their {\small{Sherwood}} run has a similar hydrodynamics and configuration as our models, whereas {\small{HVEL}} implements a more aggressive version of the energy--driven winds used in this work (including also a minimum wind speed of 600 km s$^{-1}$). Nevertheless, all their models struggle to reproduce the high \ion{C}{IV} column density absorbers. This issue is not alleviated by varying the UVB. The presence of rare strong \ion{C}{IV} absorbers in our theoretical models results from an appropriate level of resolution on the scale of the absorbers and a post-processing pipeline that closely mimics the method used by observers and accounts for individual features to calculate the column densities from Voigt profile fits to the absorption lines. The latter increases the accuracy in the estimation of the column densities.

\subsection{CIV cosmological mass density}

The evolution of the total density in \ion{C}{IV} ions with respect to the critical density is described by the comoving mass density $\Omega_{\ion{C}{IV}}$. For any ion, $\Omega_{\text{ion}}$ can be derived by summing the column densities of identified absorbers:
\begin{equation}\label{omegaion}
\Omega_{\text{ion}}(z)=\frac{H_0 m_{\text{ion}}}{c \rho_{\text{crit}}}\frac{\sum \text{N}(\text{ion},z)}{n_{\text{lov}}\Delta X},
\end{equation}
\noindent where $m_{\text{ion}}$ is the mass of the ionic species, $n_{\text{lov}}$ is the number of lines of view (lov), $\rho_{\text{crit}}$ is the critical density today and $\Delta X$ is defined in equation~\eqref{dX}.

\begin{figure}
	\includegraphics[width=\columnwidth]{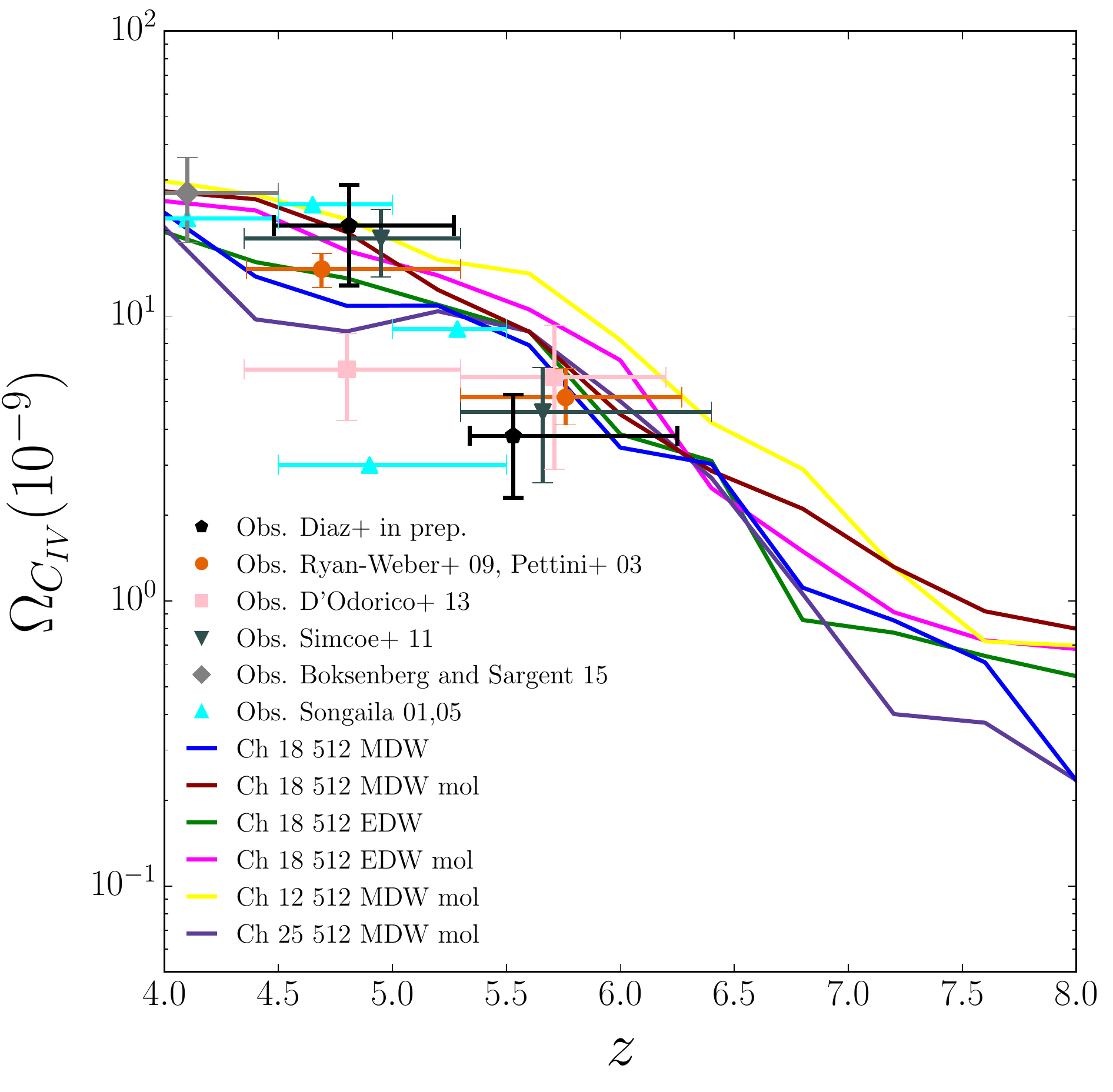}
        \caption{\ion{C}{IV} cosmological mass density at 4 $< z <$ 8. Comparison between the simulated data and observations by \citet{pettini2003} and \citet{ryanweber2009} in orange circles, \citet{songaila2001,songaila2005} in cyan triangles, \citet{simcoe2011} in dark green inverted triangles, \citet{dodorico2013} in pink squares, \citet{boksenberg2015} in grey diamond and D\'iaz et al. (in prep) in black pentagons. Pettini, Ryan-Weber and D\'iaz measurements are converted to the Planck cosmology, while for the others this recalibration was not possible due to missing details of the precise pathlength probed.}
    \label{fig:omega_c4}
\end{figure}

\noindent In the case of \ion{C}{IV}, we restricted the column densities in the sum to the range \mbox{13.8 $<$ $\log$ N$_{\ion{C}{IV}}$(cm$^{-2}$) $<$ 15.0}, to compare directly with the results of \citet{dodorico2013}. The results for all of the simulations are shown in Figure \ref{fig:omega_c4}. The theoretical results are compared to observations by \citet{pettini2003} and \citet{ryanweber2009}, \citet{songaila2001,songaila2005}, \citet{simcoe2011}, \citet{dodorico2013}, \citet{boksenberg2015} and D\'iaz et al. (in prep). It is worth mentioning that the observations by Pettini, Ryan-Weber and D\'iaz have been recalibrated to the cosmology adopted in this paper. In the other cases, missing details of the precise pathlength probed do not allow us to convert the observations to the Planck cosmology.\newline The rise of \ion{C}{IV} over the redshift period from 8 to 4 reflects both the increase in the chemical enrichment of the intergalactic medium through galactic feedback mechanisms, and the evolution of the ionization state of the IGM. At $z =$ 8, \ion{C}{IV} is largely suppressed and most of the Carbon in the box is in its neutral state. As time passes, more SN events take place and pollute the IGM with metals (as we will show in section \ref{totc}). Simultaneously, the specific intensity of the HM12 UVB around the wavelength where the \ion{C}{IV} transition occurs ($\lambda_{\ion{C}{IV}}=192$ \AA) increases by $\sim 4$ orders of magnitude between $z=$ 8 and 4. At all redshifts, each simulation is broadly consistent with current observational data. \newline Using results from a hybrid wind model, \citet{finlator2015,finlator2016} integrated a power law fit to the column density distribution function to calculate the evolution of the comoving mass density of \ion{C}{IV}. Instead, we calculate $\Omega_{\ion{C}{IV}}$ by summing the column density of the absorbers using equation~\eqref{omegaion}. This method is more accurate when there are uncertainties in the slope of the \ion{C}{IV}--CDDF, especially at high column densities.

\begin{figure*}
	\includegraphics[scale=0.30]{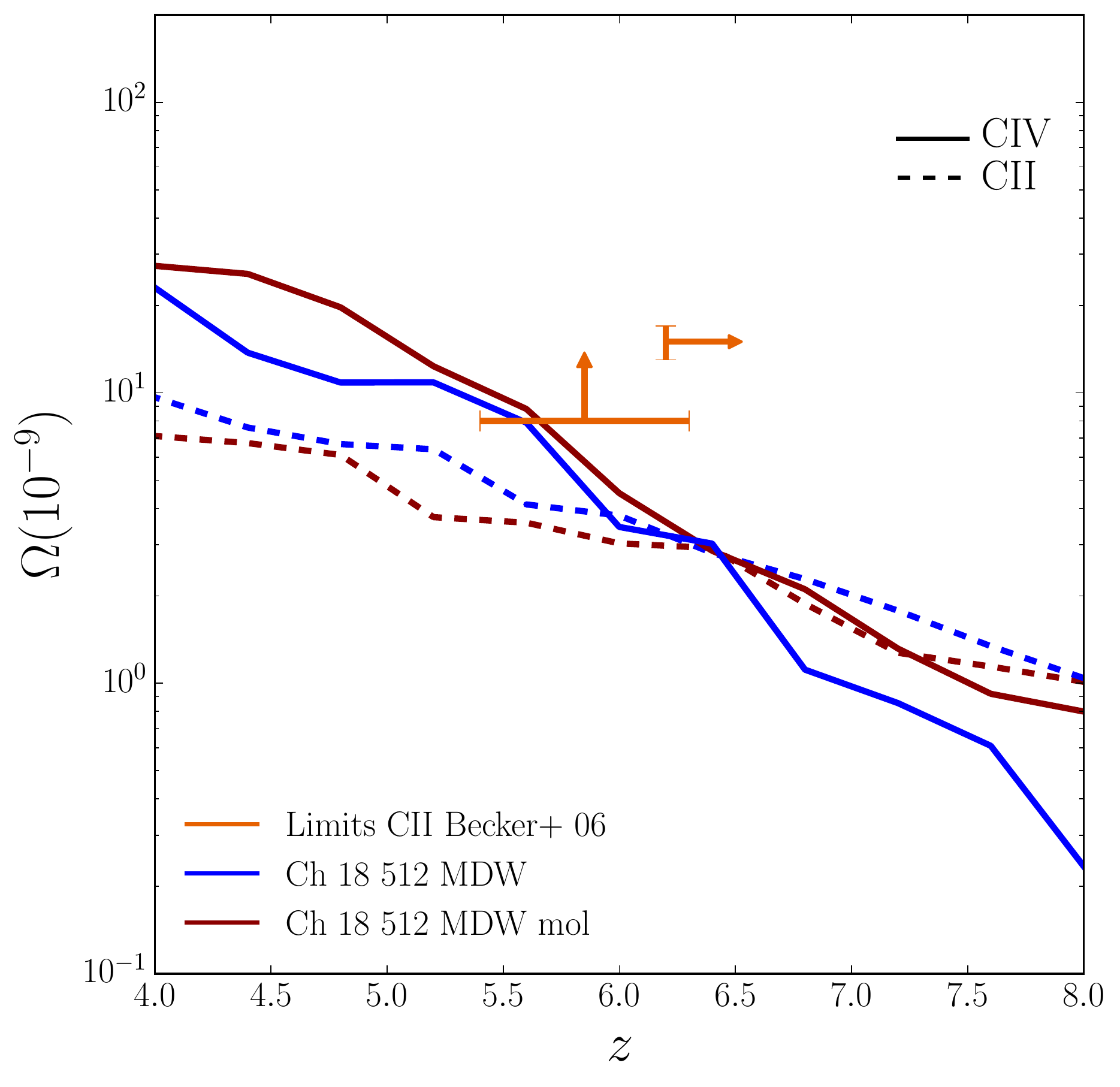}
	\includegraphics[scale=0.30]{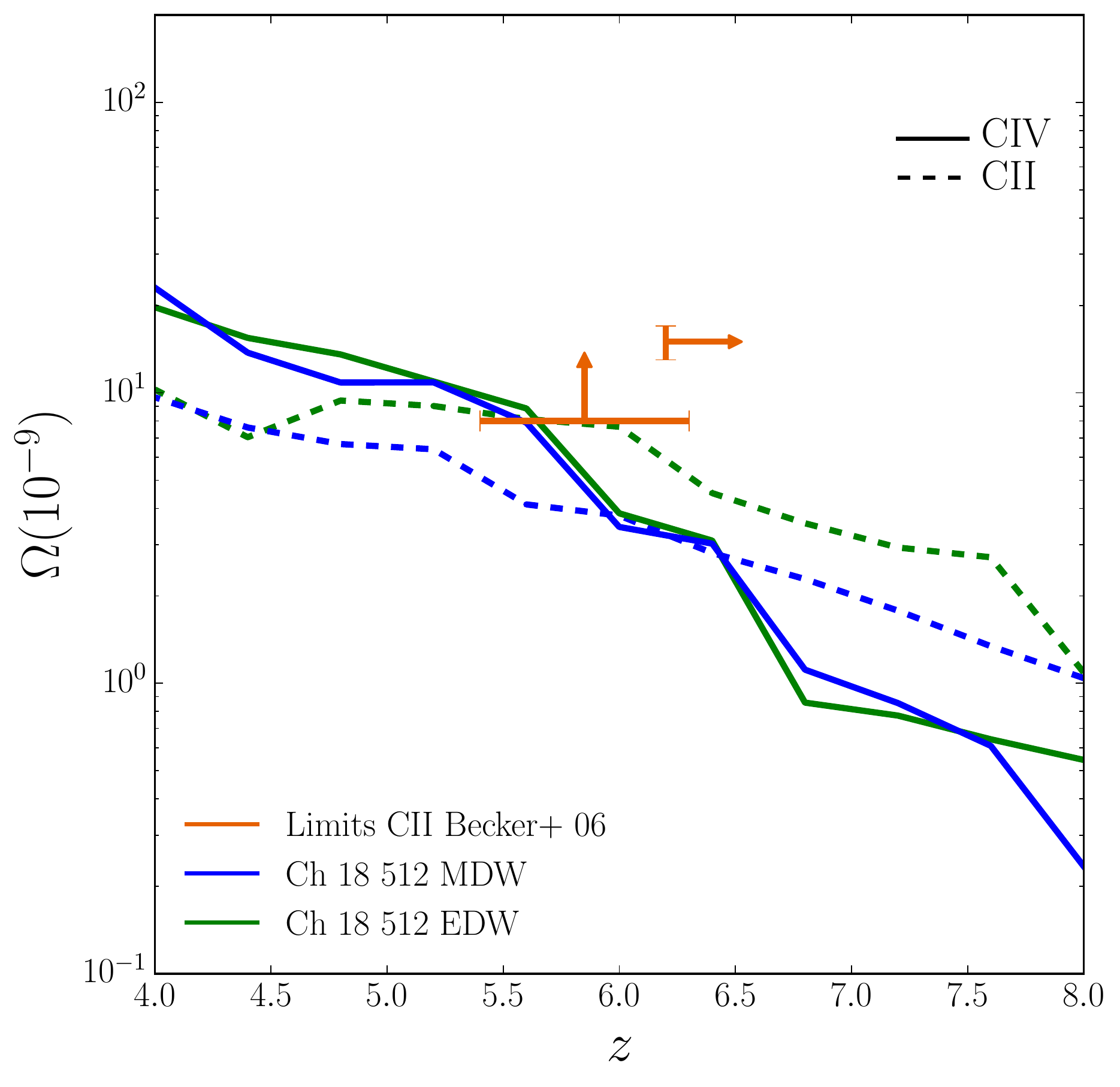}
	\includegraphics[scale=0.30]{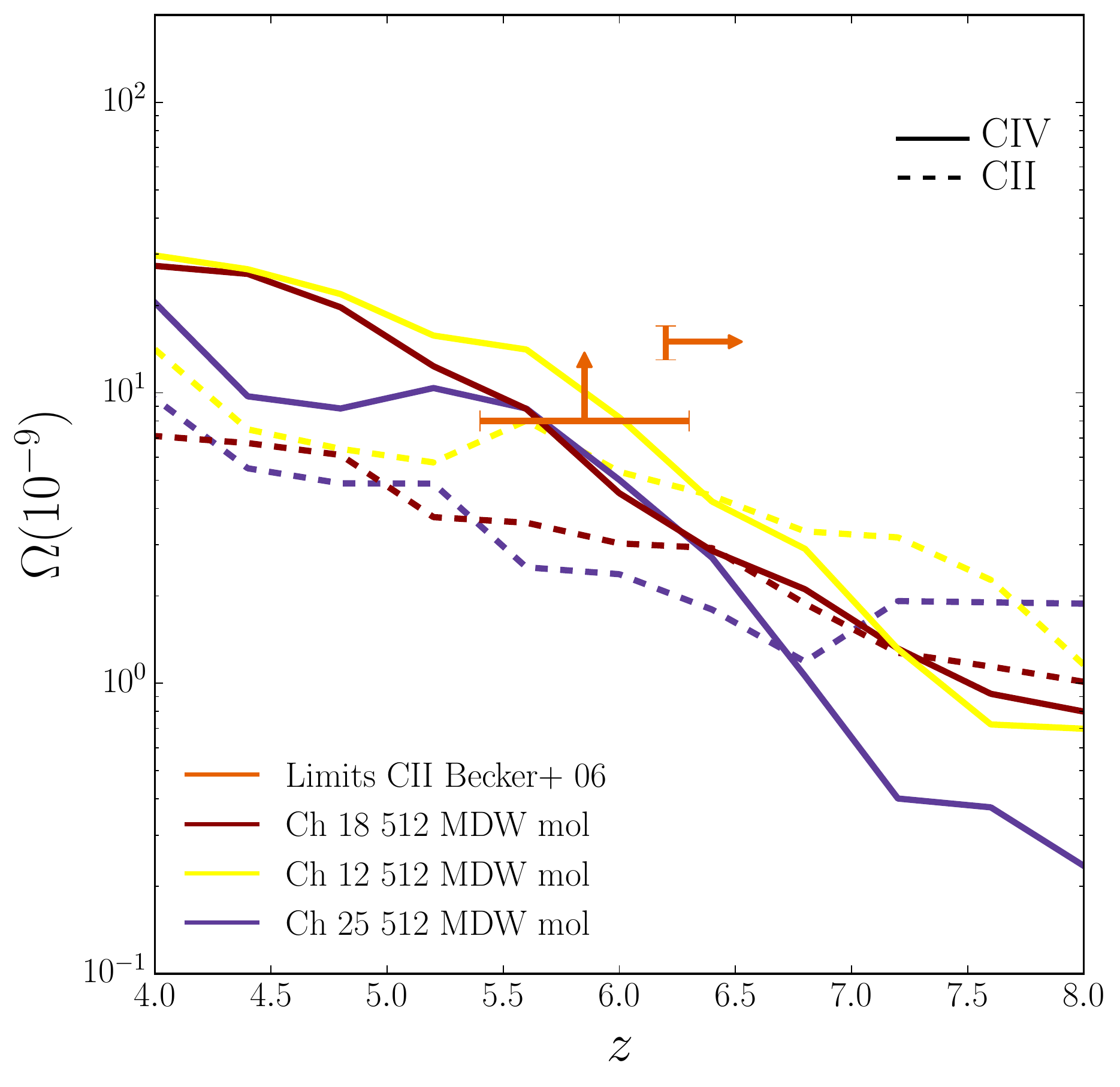}
        \caption{Evolution of the \ion{C}{II} and \ion{C}{IV} cosmological mass density. The left and central panel compare simulations with 18 cMpc/$h$ box size. On the left is displayed the case of simulations with MDW feedback with and without low temperature metal and molecular cooling included. In the middle, we show the configuration without low-T metal and molecular cooling for the runs with MDW and EDW feedback. On the right hand side, a resolution and box size test with the following models is shown: Ch 12 512 MDW mol, Ch 18 512 MDW mol and Ch 25 512 MDW mol. In all the panels, the solid lines show the evolution of $\Omega_{\ion{C}{IV}}$ for 13.8 $<$ $\log$ N$_{\ion{C}{IV}}$(cm$^{-2}$) $<$ 15.0, and the dashed lines $\Omega_{\ion{C}{II}}$ in the range 13.0 $<$ $\log$ N$_{\ion{C}{II}}$(cm$^{-2}$) $<$ 15.0. The orange points with errors represent the observational lower limits for $\Omega_{\ion{C}{II}}$ from \citet{becker2006}.}
    \label{fig:omega_c2}
\end{figure*}

\subsection{CII cosmological mass density}

We extend the analysis to \ion{C}{II} and compare with the observational lower limits of \citet{becker2006} in Figure~\ref{fig:omega_c2}, taking into account systems whose column densities are in the range 13.0 $<$ $\log$ N$_{\ion{C}{II}}$(cm$^{-2}$) $<$ 15.0. In the context of Reionization, it is interesting to consider the evolution of \ion{C}{II} (a low ionization state of Carbon with ionization energy of 11.26 eV) with respect to \ion{C}{IV}, whose ionization energy is more than 4 times larger (47.89 eV). Since the UVB is softer at earlier times, \ion{C}{II} is expected to be more dominant than higher ionization states of Carbon at high redshifts.\newline On the left and central panels of Figure~\ref{fig:omega_c2} we draw a comparison between different physical scenarios among the simulations with boxsize 18 cMpc/$h$: the further left relates simulations with the same feedback prescription (MDW) with and without low temperature metal and molecular cooling. The run with low-T cooling included produces slightly more \ion{C}{IV} (and less \ion{C}{II}) than the fiducial run, but the difference is only significant at $z>$ 6.5. In the middle panel, we contrast simulations without low temperature metal and molecular cooling with different feedback prescriptions: MDW and EDW. The different physical scenarios do not have a strong impact on the evolution of \ion{C}{IV}, as seen also in Figure~\ref{fig:omega_c4}. The trends for \ion{C}{II} are more dependent on the model, because this low--ionization state is very sensitive to the densities of the absorbers (which are correlated to the strength of the feedback prescription), and the Ch 18 512 EDW run is in good agreement with the observations of \citet{becker2006}. \newline Although the run with lowest resolution, Ch 25 512 MDW mol, produces slightly degraded results, the evolution of $\Omega_{\ion{C}{IV}}$ does not change significantly with the spatial resolution or size of the simulations (see the rightmost panel of Figure~\ref{fig:omega_c2}). \newline On the other hand, \ion{C}{II} trends are more influenced by the resolution. Ch\- 12\- 512\- MDW\- mol better reproduces the observed evolution of this ion. The absorption features are affected by the box size (and the comoving softening) for different reasons: the absorption path calculated to create the synthetic spectra grows with the box size, as well as the size of the pixels, while the definition of the individual features decreases for larger configurations. The Ch 12 512 MDW mol run is in better agreement with the observational limits for \ion{C}{II} at high redshift, indicating that higher resolution in the simulations helps to better describe the absorption features, in particular for low ionization states.\newline

The most interesting part of the analysis arises when the \ion{C}{II} and \ion{C}{IV} curves are compared for each model: in the fiducial simulation, the amount of \ion{C}{IV} exceeds the amount of \ion{C}{II} at $z \sim$ 6, consistent with the fact that the Universe is approaching the tail of the EoR. This result is seen in all of the simulations, with a crossover in the range of $z \sim$ 6--6.5.\newline Despite the fact that in some runs the synthetic \ion{C}{II} does not reach the lower limits predicted in \citet{becker2006}, the crossover of these ions at $z \sim$ 6--7 is very promising. Other theoretical works that also follow the trend of these ions \citep[e.g.][]{finlator2015} obtained a crossover between $\Omega_{\ion{C}{II}}$ and $\Omega_{\ion{C}{IV}}$ at $z \sim$ 8. However, a later crossover of \ion{C}{II} and \ion{C}{IV} at $z \sim$ 6 is more consistent with the observations and the current paradigm of the tail of Reionization.

% As a final remark, we warn the reader that the evolution of Hydrogen Reionization, is regulated by the UVB in the simulations. An evolving uniform UVB ionizes low--density regions in post-processing faster than the dense regions near galaxies. In addition, the simulated boxes are rapidly ionised with this method, not slowly as the average UVB would imply. When calculating the evolution of the cosmological mass densities of the ions we are mainly constraining the progression of a uniform UV background inside of a region of the Universe that has been enriched at $z \sim$ 6. Nonetheless, the agreement between the observations and the mock catalog of column densities is good. Our conclusions on the likely end of the EoR are based on the assumption that our results hold when extrapolated to larger volumes.\newline

\subsection{Total Carbon cosmological mass density}
\label{totc}

\begin{figure}
	\includegraphics[width=\columnwidth]{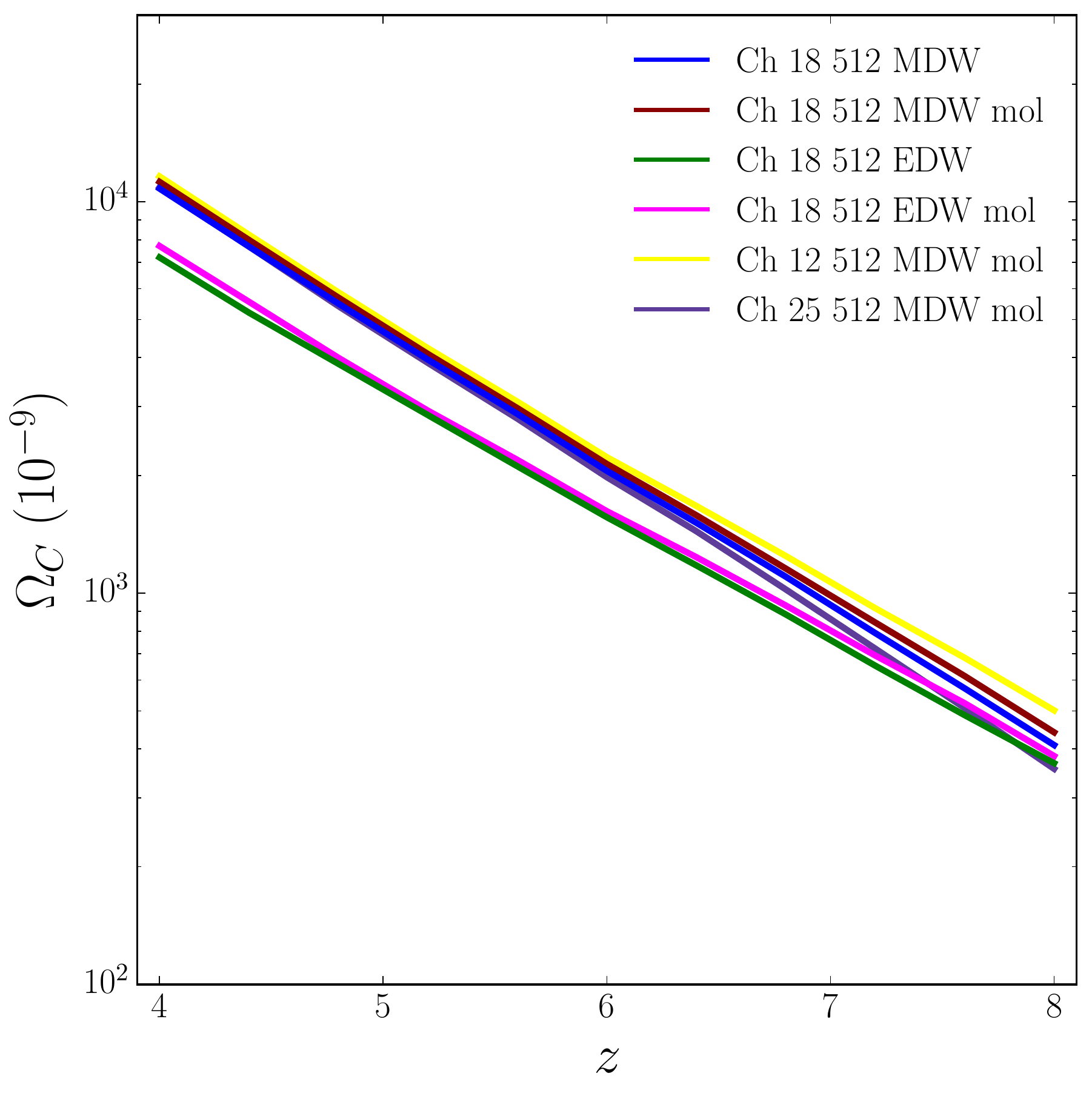}
        \caption{Evolution of the total Carbon cosmological mass density for our simulations. This is a purely theoretical prediction, since $\Omega_{\text{C}}$ is not (directly or indirectly) observable.}
    \label{fig:omega_c}
\end{figure}

To conclude section \ref{civ}, in Figure~\ref{fig:omega_c} we analyse the evolution of the cosmological mass density of (total) Carbon in the simulated boxes, $\Omega_{\text{C}}$. This is calculated differently than $\Omega_{\ion{C}{IV}}$ and $\Omega_{\ion{C}{II}}$: here we sum the amount of Carbon in each gas particle in the simulation and divide by the comoving volume. Thus, $\Omega_{\text{C}}$ gives an estimation of the total amount of Carbon at high redshift. The trend is consistent with the enrichment history of the Universe, increasing by 1.5 orders of magnitude from $z =$ 8 to 4. At $z=4$, our results are compatible with those of \citet{tescari2011}. Given the different methods used to calculate $\Omega_{\text{C}}$ and $\Omega_{\ion{C}{IV}}$ \& $\Omega_{\ion{C}{II}}$, the first quantity should not be directly compared with the other two. Moreover, the evolution of the total Carbon cosmological mass density is a purely theoretical prediction of our simulations, since $\Omega_{\text{C}}$ is not (directly or indirectly) observable. \newline The overall evolution of C is nearly independent on the nature of the cooling (blue vs red and green vs magenta lines), but depends at low redshift on the feedback mechanism. The EDW models quench the formation of stars in galaxies more effectively than MDW, therefore less Carbon is produced and $\Omega_{\text{C}}$ is lower.\newline Figure~\ref{fig:omega_c} also compares simulations with different spatial resolutions (the comoving softening of the different configurations scales with the size of the box). The Ch 12 512 MDW mol, Ch 18 512 MDW mol and Ch 25 512 MDW mol runs have comoving softening of 1.0, 1.5 and 2.0 ckpc/$h$ (yellow, red and purple lines), respectively. These simulations predict a similar trend for the mass density of Carbon and at low redshift converge to the same $\Omega_{\text{C}}$. However, at high redshift the resolution affects the amount of metals. The run with the highest resolution (Ch 12 512 MDW mol, yellow line) can resolve higher densities at earlier times and therefore better describes star formation in galaxies and produces more Carbon than Ch 18 512 MDW mol (red) and, particularly, Ch 25 512 MDW mol (purple).\newline

\vspace{0.0cm} \begin{figure*} \centering \includegraphics[scale=0.35]{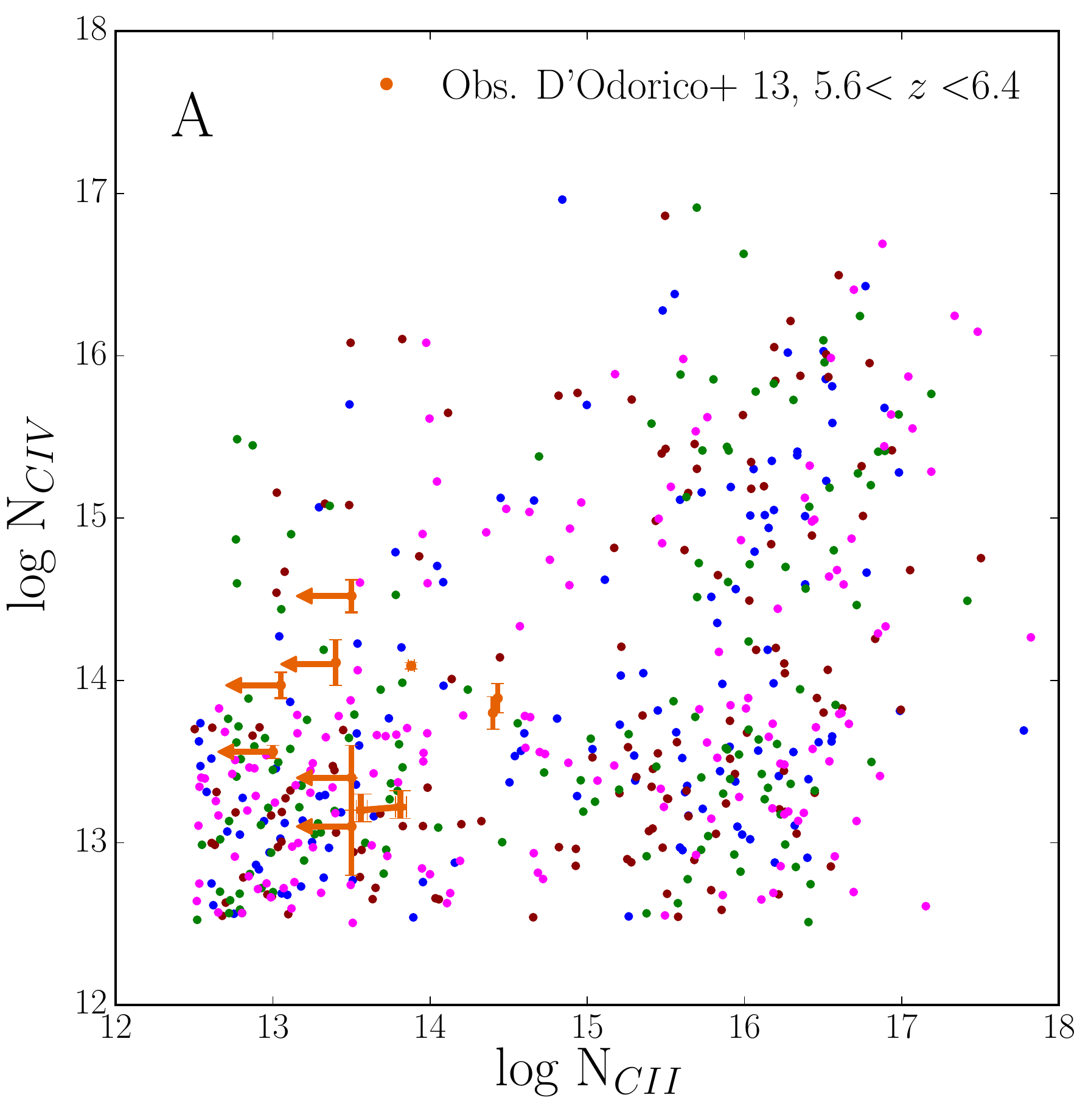} \includegraphics[scale=0.35]{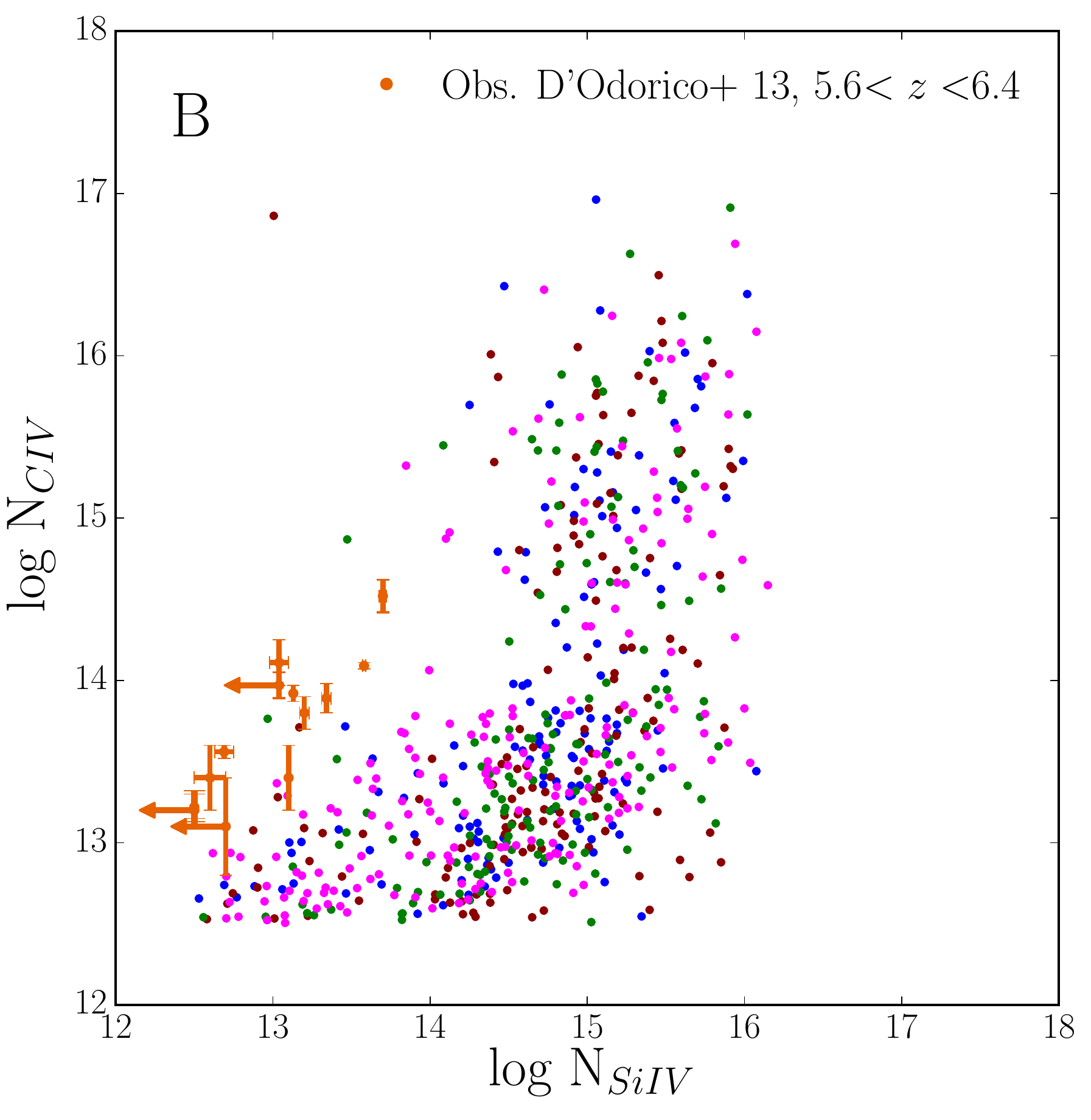} \vspace{0.0cm} \hspace{0.0cm} \includegraphics[scale=0.35]{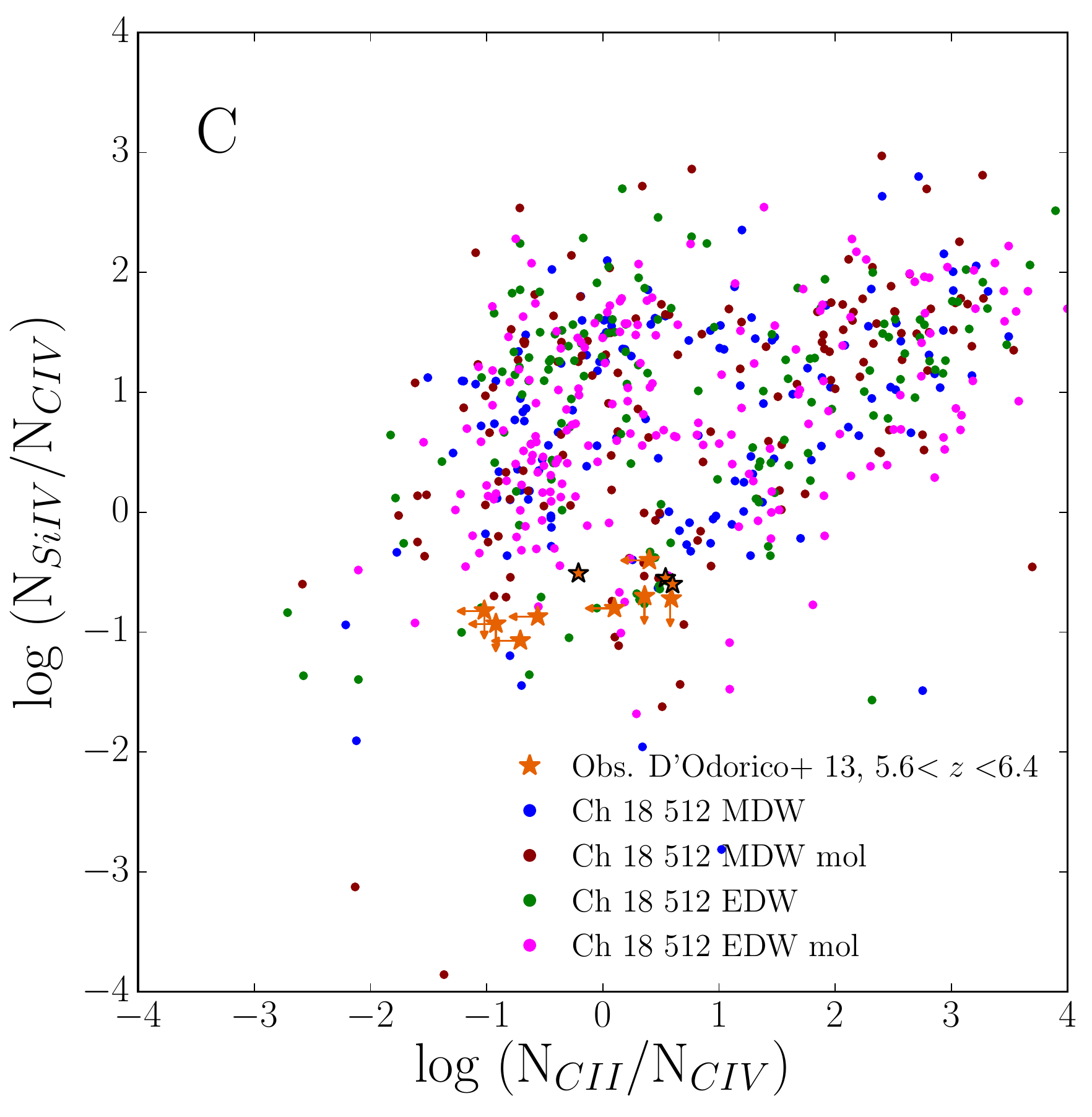} \includegraphics[scale=0.35]{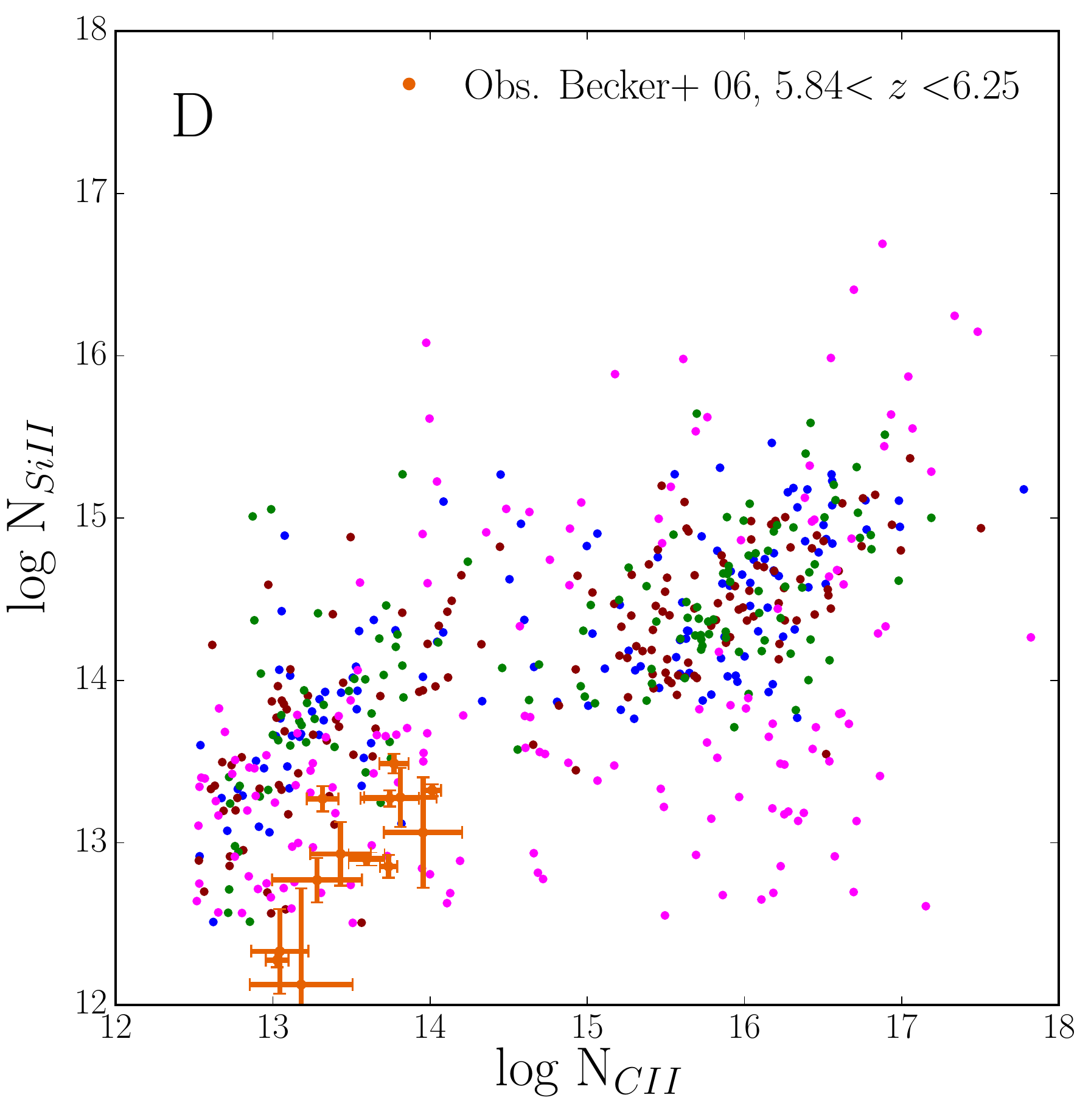} \vspace{0.0cm} \hspace{0.0cm} \includegraphics[scale=0.35]{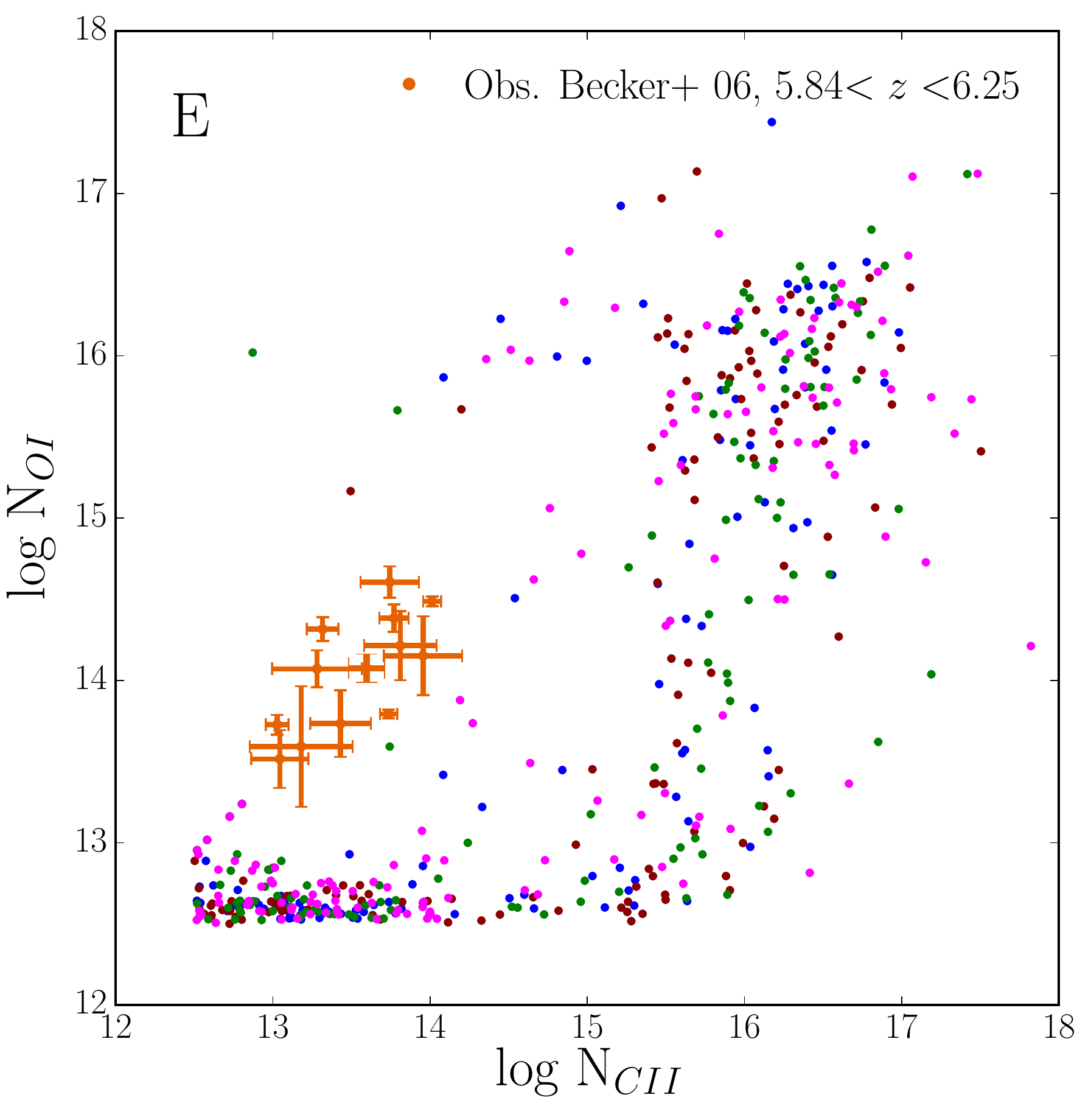} \includegraphics[scale=0.35]{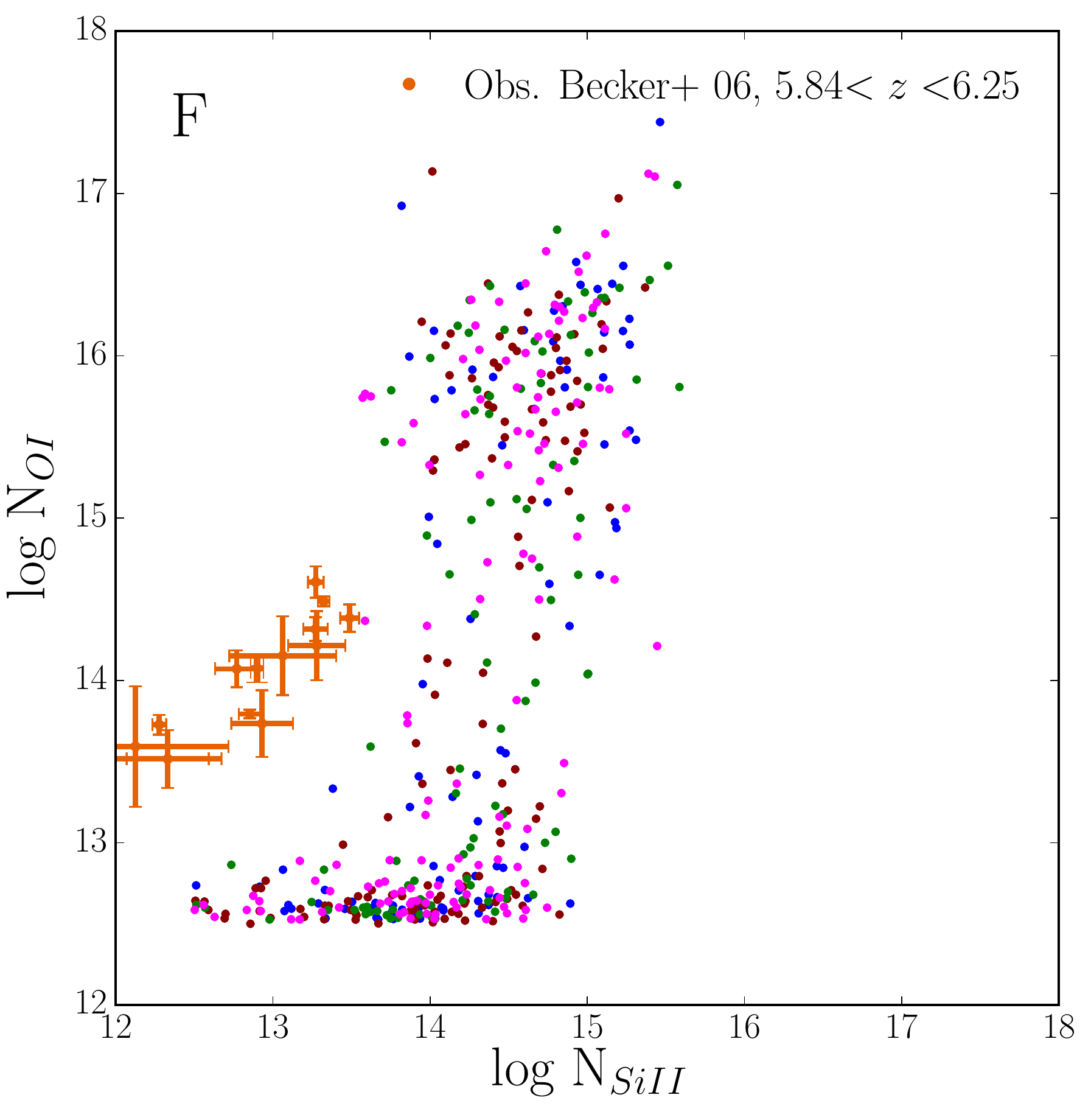} \vspace{0.0cm} \caption{Column density relationships among metal absorption lines at $z =$ 6. Panels A, B and C compare with systems observed by \citet{dodorico2013} in the redshift range of (5.6--6.4) and D, E and F with absorbers detected by \citet{becker2006} in the redshift range of (5.84--6.25), both in orange. From the synthetic spectra of each ions, the column densities are calculated and displayed for the simulations with box size of 18 Mpc/$h$ (the label keys are shown in the panel C), but there is not a remarkable difference between the different prescriptions (feedback and cooling) imposed in each case. The simulated data displayed are above the threshold noise at $\log$ N$_{\text{th}} =$ 12.5 (cm$^{-2}$). This selection criterion results in an effective number of synthetic systems of $\sim$18\% from the initial 1000 random lines of sight traced. In panel C, the observations by \citet{dodorico2013} produced 3 systems with column densities reported (represented by the black stars) and 8 with upper limits in N$_{\ion{C}{II}}$ and/or N$_{\ion{Si}{IV}}$, represented by arrows.}  \label{fig:correlations} \end{figure*}

\section{Column density relationships of metal absorption lines}
\label{corr}

One of the advantages of the numerical approach is that it reproduces a large number of sightlines, improving the statistical estimation of the column densities at redshifts that are hard to access with current observations. By convolving our synthetic spectra with Gaussian noise and extracting column densities using Voigt profile fitting, our simulated results closely mimic the observed spectra.\newline As mentioned in section~\ref{c4cddf}, this procedure produces a very large number of systems with low column densities, that display non physical linear correlations among the metal absorption features. To avoid this issue, a cut--off in column density, $\log$ N$_{\text{th}}$ (cm$^{-2}$) $\ge$ 12.5, has been imposed to the synthetic data, and systems below this noise threshold have not been taken into account in the statistics.\newline

We explore relationships in column densities among di\-ffe\-rent ionic species in a uniform UV background HM12 at $z =$ 6 using random lines of sight to emulate the observational method and compare with the observations available to date in Figure~\ref{fig:correlations}. In panel A, N$_{\ion{C}{II}}$ vs. N$_{\ion{C}{IV}}$ are compared, panel B displays N$_{\ion{Si}{IV}}$ vs. N$_{\ion{C}{IV}}$ and panel C the ratio of N$_{\ion{C}{II}}$/N$_{\ion{C}{IV}}$ vs. N$_{\ion{Si}{IV}}$/N$_{\ion{C}{IV}}$. All these quantities are compared to systems detected by \citet{dodorico2013}, while panels D, E and F draw a comparison with low ionization states observed by \citet{becker2006}. The comparisons N$_{\ion{C}{II}}$ vs. N$_{\ion{Si}{II}}$, N$_{\ion{C}{II}}$ vs. N$_{\ion{O}{I}}$ and N$_{\ion{Si}{II}}$ vs. N$_{\ion{O}{I}}$ are shown in panels D, E and F, respectively. In all cases, we display the simulations with 18 cMpc/$h$ box size, to analyze possible variations among the predicted values with different prescriptions of feedback and cooling. However, the dispersion of the data points do not allow clear distinctions between the models.\newline The high ionization states are relatively well represented by the simulated data with the UV ionizing background implemented, especially N$_{\ion{C}{IV}}$ as seen in panels A and C. On the other hand, the low ionization species are produced far in excess by the simulations and in all the different configurations considered the column densities are overestimated for \ion{O}{I} and \ion{Si}{II}. \newline \ion{C}{II} is a very interesting case in this study. The numerical estimation of \ion{C}{II} column densities do not seem to be well reproduced by the simulations, although in most of the cases the observational data are just a lower limit, and a bimodality in the distribution of column densities appears, which will be analyzed in section \ref{bimodality}.\newline The scatter plots in many cases struggle to reproduce the observed column densities, regardless of the feedback model used, possibly due to the UV background implemented. The parameters that characterize HM12 can suppress the low ionization states. \citet{bolton2011} claimed that N$_{\ion{Si}{IV}}/$N$_{\ion{C}{IV}}$ is sensitive to the spectral shape of the UV background, whereas \citet{finlator2015} showed that N$_{\ion{C}{II}}/$N$_{\ion{C}{IV}}$ is sensitive to the overall intensity of the UV background and its normalization. We are currently studying the effect of variations in the normalization and hardness of the HM12 UVB. The results will be presented in a follow-up paper.\newline \ion{O}{I} receives special attention because its ionization energy is similar to the one of \ion{H}{I}.  Besides the overprediction of N$_{\ion{O}{I}}$ in our simulations\footnote{There is also a constant trend of N$_{\ion{O}{I}}$ at low column densities, possibly indicating that in low-density environments Oxygen is mostly found in higher ionization states due to the effect of the UVB.}, the large column densities in this ionic species may indicate that there is a huge amount of \ion{O}{I} self-shielded that it is not being detected due to the low probability of tracing a high--density region with the current observational methods. It is worth noting that, to date, there is not a self-consistent scheme in the simulation literature that accounts for the self-shielding of the ions (corrections for self-shielding at $z <$ 5 can be found in \citealt{bird2014} and \citealt{bolton2016}). This could make the estimated column densities of the low ionization states less accurate.

\subsection{Bimodality in CII column densities}
\label{bimodality}

\begin{figure}
	\includegraphics[width=\columnwidth]{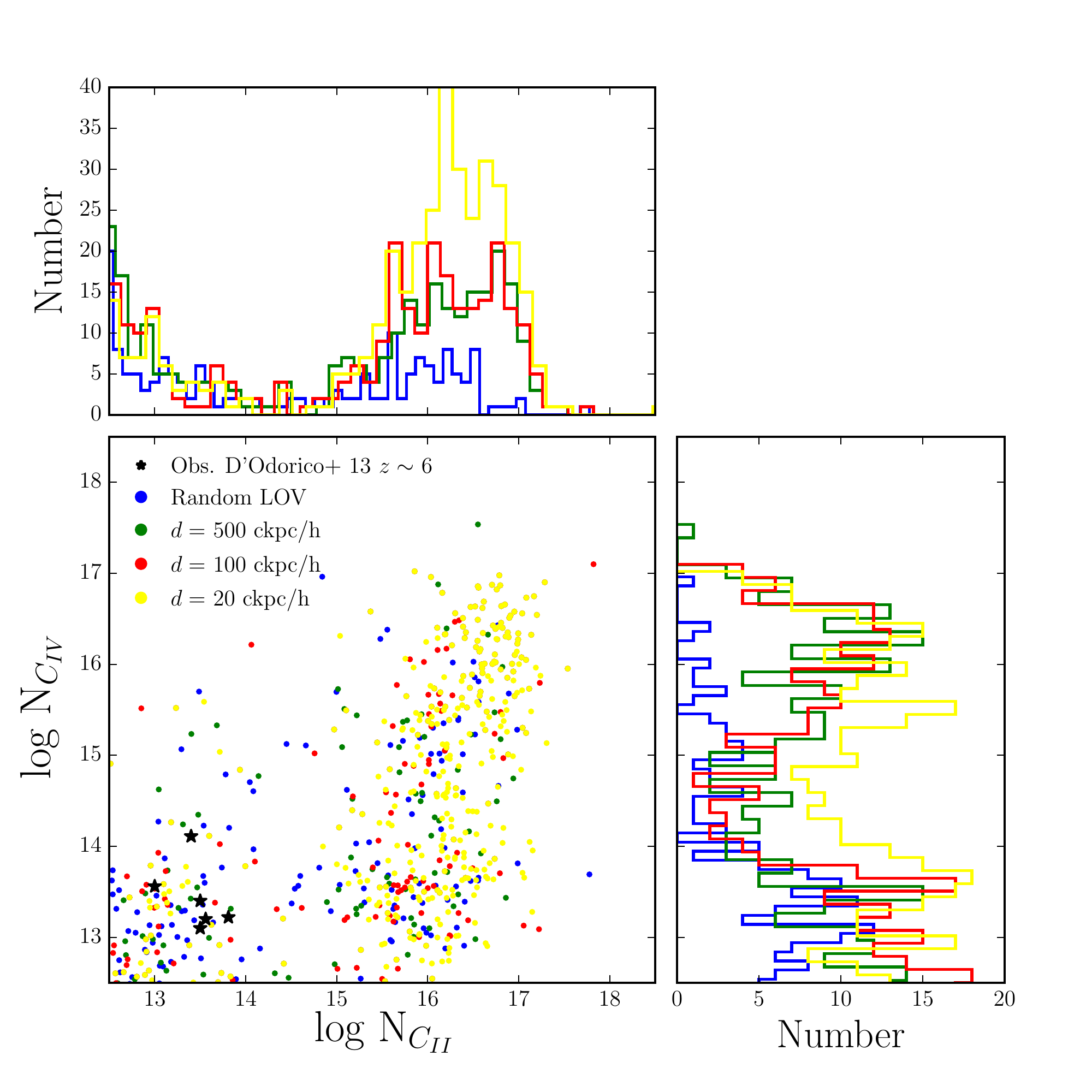}
        \caption{\ion{C}{II} vs \ion{C}{IV} column densities in different environments at $z =$ 6 for the reference simulation, Ch 18 512 MDW. Using different impact parameters from the centre of the galaxies it is possible to understand what drives a bimodality in N$_{\ion{C}{II}}$. This low ionization state of C is mostly found in the CGM, but there are some traces of it in the IGM and, so far, it has been observed in the latter, with low column densities, as it is shown in the plot (the black stars are observational data at $z=$ 6 from \citealt{dodorico2013}). The blue points are theoretical predictions using random lines of sight (cf. panel A in Figure~\ref{fig:correlations}). When the sightlines are chosen to be close to the centre of the galaxies, the distribution in \ion{C}{II} is shifted to larger column densities, indicating that high-$z$ galaxies are huge reservoirs of \ion{C}{II}, not detected yet due to the low likelihood of reaching these overdense regions with QSOs lines of sight.}
    \label{fig:histo_c}
\end{figure}

The location of the low-ionization states of metals at $z \sim$ 3, for instance \ion{C}{II} and \ion{O}{I}, traces the regions close to the centre of the galaxies or DLA systems. This correlation of gas ionization state and proximity to the centre of the nearest galaxy may no longer be valid at high redshift, as \ion{C}{II} may be tracing either DLAs or the IGM at low temperature.\newline Observations at $z \sim$ 6 exhibit N$_{\ion{C}{II}} \leq$ 10$^{14}$ cm$^{-2}$, while at higher redshift only lower limits in \ion{C}{II} are available. In all the simulations, we found a remarkable bimodal distribution in the N$_{\ion{C}{II}}$ calculated using random sightlines through each box (see e.g. panel A in Figure~\ref{fig:correlations}). In order to test the nature of this bimodality, in Figure~\ref{fig:histo_c} we plot the column densities of \ion{C}{II} and \ion{C}{IV}, using lines of sight at different impact parameters from the centre of the galaxies in the fiducial run Ch 18 512 MDW ($d =$ 20, 100 and 500 ckpc/$h$ in yellow, red and green points, respectively). At each $d$, we extracted a thousand lov around halos with masses in the range 10$^{9-10} M_{\sun}$/$h$. The blue points represent the column densities along 1000 random lines of sight through the box shown in panel A of Figure~\ref{fig:correlations}. The black stars are observational data from \citet{dodorico2013}. The top panel in Figure~\ref{fig:histo_c} displays the distribution of the \ion{C}{II} absorption features and the right--most window the corresponding distribution of \ion{C}{IV} column densities above the noise threshold.

\noindent When the impact parameter $d$ approaches the galaxies, the peak at high column densities becomes sharper, indicating that there are more \ion{C}{II} absorbers in the CGM where the temperature and density are larger than in the IGM. For $d =$ 20 ckpc/$h$, $\sim$ 3 pkpc at this redshift, the absorbers are located inside or in the outskirts of the galaxies (spectroscopic confirmation of galaxies at $z =$ 6--8 with the Hubble Space Telescope shows that the radii of these galaxies are in the range of 0.6--1.1 pkpc, \citealt{jiang2013}). Absorbers are gravitationally bounded to the halos in the simulated sample, which have virial radii of 3--30 pkpc.\newline The distribution of the \ion{C}{II} column densities at $d =$ 20 ckpc/$h$ indicates that the few observations towards high-redshift QSOs currently available have not yet detected rare systems with large column densities in \ion{C}{II}, due to the very low likelihood of intersecting a galaxy at a low impact parameter (that would produce a deep absorption feature). Furthermore, \citet{dodorico2013} have only identified \ion{C}{II} absorptions that were first detected as \ion{C}{IV} doublets. Thus, any low ionization gas that may have produced a strong \ion{C}{II} absorber and no associated \ion{C}{IV} has not been reported. However, future observations of QSOs will raise the number of metal absorption lines detected and cover a larger range in column densities. \newline The results for N$_{\ion{C}{II}}$ with random lines of sight (blue line) agree with numerical results by \citet{oppenheimer2009}, \citet{keating2016} and a photoionization comparison by \citet{dodorico2013}, predicting that \ion{C}{II} lies mostly in overdense regions with $\delta = \frac{\rho_{\text{gas}}}{\langle \rho \rangle}$ - 1 $ >$ 10, and there are some traces in low temperature regions (mostly the IGM probed by current observations) where its column densities are of the same order of N$_{\ion{C}{IV}}$.\newline Finally, numerical results in the CGM (yellow and red cases in Figure~\ref{fig:histo_c}) lead to a bimodality in \ion{C}{IV} and a clear excess of high column density systems at low impact parameters. \citet{oppenheimer2009} showed that the properties of the \ion{C}{IV} absorbers depend mostly on the evolution of the environment of their hosting galaxy (distance, mass and metallicity). By tracing the surroundings of the galaxies in halos with masses 10$^{9-10} M_{\sun}$/$h$, we confirm that there is a correlation between \ion{C}{IV} column densities and the proximity of the galaxies that host the absorbers.\newline
%\textbf{On the other hand, it is worth mentioning that when tracing a line of sight with a given impact parameter from the centre of a halo there is a possibility of encounter/intercept another halo in the box. However, the bimodal distribution is not driven by this effect: tests with very large impact parameters should reproduce the results obtained with random lines of sight. This is confirmed by comparing the green and blue histograms ($d =$ 500 ckpc/$h$ and lines of view schocastically selected). These results significantly differ from tests with smaller impact parameters $d$, thus the bimodality in the distribution of N$_{\ion{C}{II}}$ is intrinsic to the data.}

\section{Cosmological mass density of HI at redshift lower than 6}
\label{hi}

\begin{figure}
	\includegraphics[scale=0.45]{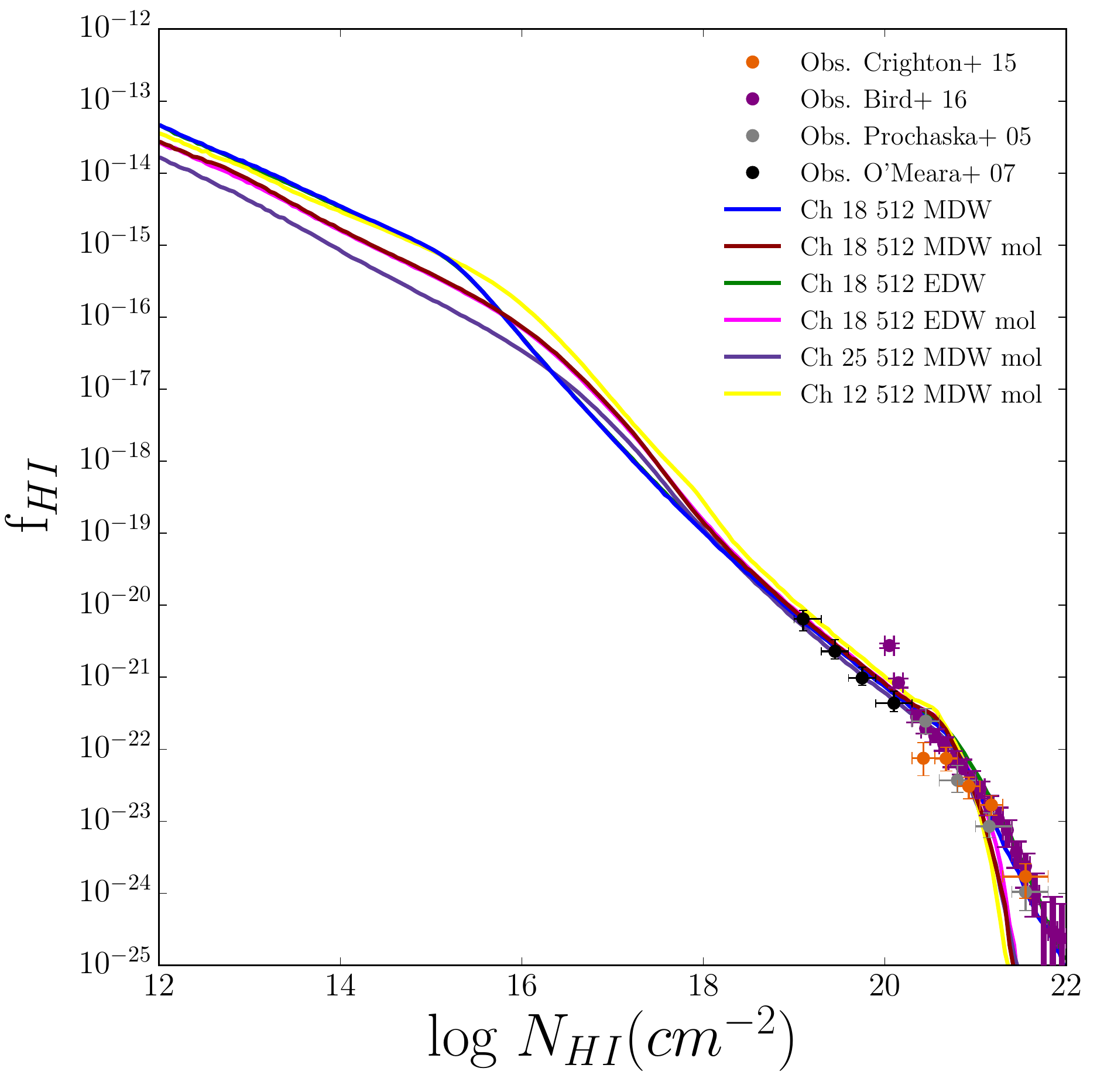}
        \caption{\ion{H}{I} column density distribution function at $z=$ 4 in the range 12 $<$ $\log$ N$_{\ion{H}{I}}$ (cm$^{-2}$) $<$ 22 for all the simulations described in Table \ref{table_sims} and comparison to observations by \citet{prochaska2005} in grey, \citet{omeara2007} in black, \citet{crighton2015} in orange and \citet{bird2016} in purple.}
    \label{fig:HICDDF1}
\end{figure}

\begin{figure}
	\includegraphics[scale=0.45]{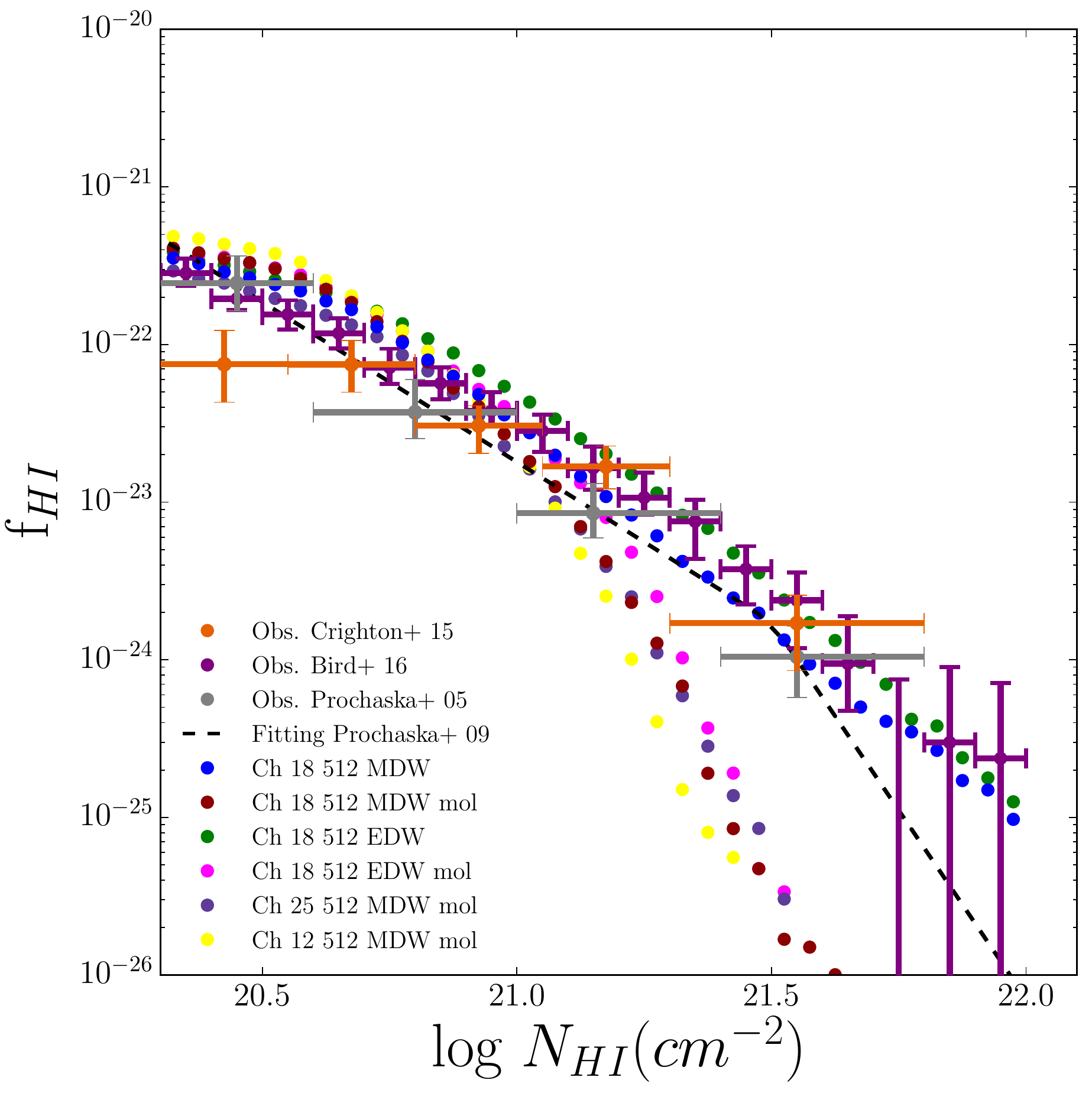}
        \caption{Same as Figure \ref{fig:HICDDF1} but zoomed in the DLA range: 20.3 $<$ $\log$ N$_{\ion{H}{I}}$ (cm$^{-2}$) $<$ 22. The black dashed line represents the fitting function by \citet{prochaska2009} for DLA systems at redshift 4.0--5.5.}
    \label{fig:HICDDF2}
\end{figure}

The estimation of the amount of neutral Hydrogen can be derived using DLAs. It is worth noting that at high redshift ($z >$ 5) the nature of DLAs is ambiguous \citep[e.g.][]{simcoe2012}, since the definition of a DLA is by column density only (N$_{\ion{H}{I}} >$ 10$^{20.3}$ cm$^{-2}$) and does not discriminate between dense pockets of neutral Hydrogen in the IGM, where Reionization is still progressing, and DLAs associated with self-shielded gas in galaxies (like at $z <$ 5). \newline
However, it is possible to simulate DLA systems to calculate their corresponding comoving mass density $\Omega_{\text{DLA}}$ and from it, estimate $\Omega_{\ion{H}{I}}$. We identify the halos with masses above 10$^9 M_{\sun}/h$ (using a halo finder routine, the halos are sorted by mass) and extract cubes of size 200 ckpc/h around the center of mass of one thousands of them (randomly distributed in mass). In this way, we guarantee that the simulated DLAs are not affected by a selection bias in density. Then, the neutral Hydrogen density in the cube is integrated along the line of sight and projected into a 2D grid of N$_{\ion{H}{I}}$. With these column densities, we finally calculate the \ion{H}{I}--CDDF, $f_{\text{\ion{H}{I}}}(\text{N},X)$, as:\newline
\begin{equation}\label{HI_CDDF}
f_{\ion{H}{I}}(\text{N},X)=\frac{n_{\text{sys}}(\text{N}, \text{N}+\Delta \text{N})}{\Delta X \cdot n_{\text{cube}} \cdot n_{\text{grid}}^2},
\end{equation}
\noindent where $n_{\text{sys}}$ is the number of systems in the interval $(\text{N}, \text{N}+\Delta \text{N})$, $\Delta X$ the absorption path defined in equation~\eqref{dX}, $n_{\text{cube}}$ the number of cubes generated and $n_{\text{grid}}$ the number of bins in the grid. The parameters selected to achieve a good resolution of the systems that contain neutral Hydrogen are:   $n_{\text{cube}} =$ 1000 and $n_{\text{grid}} =$ 64 (i.e. the grid size is 3.1 ckpc/h).\newline Figure~\ref{fig:HICDDF1} shows the \ion{H}{I}--CDDF, with column densities in the range 12 $<$ $\log$ N$_{\ion{H}{I}}$(cm$^{-2}$) $<$ 22, for all the simulations at $z=$ 4 and compares with the observations of \citet{prochaska2005}, \citet{crighton2015}, \citet{bird2016} (all in the column densities corresponding to DLA systems at $z \geq$ 4) and \citet{omeara2007} in the super Lyman limit systems (LLS) range at $z \sim$ 3.4\footnote{Although these systems are at $z < 4$, they are a reliable benchmark for our simulated data, since there is evidence of low to no evolution in this redshift range.}. Note that there is a very large number of systems with low column densities in all the models and DLAs are extremely rare. The agreement with the observational data is good, indicating that the method implemented to find neutral systems works well at these redshifts.\newline Simulations with smaller comoving softening (or smaller box size), better resolve the Ly$\alpha$ forest.  In Figure~\ref{fig:HICDDF1} the yellow line (simulation Ch 12 512 MDW mol) is above the red line, corresponding to the run Ch 18 512 MDW mol, and the latter is above the purple line (Ch 25 512 MDW mol, the run with the lowest resolution in the suite of simulations) at $\log$ N$_{\ion{H}{I}}$(cm$^{-2}$) $<$ 19.\newline At DLA column densities, the distributions follow a double--power law, as discussed by \citet{zwaan2006} and \citet{prochaska2009}, with a knee at $\log$ N$_{\ion{H}{I}}$ (cm$^{-2}$) $\sim$ 21 in all of the cases considered. Also, at N$_{\ion{H}{I}} >$ 10$^{21}$ cm$^{-2}$, where molecular cooling plays an important role in the star forming regions, the \ion{H}{I}--CDDF in models with molecular cooling implemented drops off more rapidly than in the other models.\newline In Figure~\ref{fig:HICDDF2} a zoom of the DLA region is shown ($\log$ N$_{\ion{H}{I}}$(cm$^{-2}$) in the range $[$20.3 , 22.0$]$). Overplotted are the most recent observational data of \ion{H}{I} absorbers at high redshift by \citet{prochaska2005}, \citet{crighton2015}, \citet{bird2016} and the fitting function proposed by \citet{prochaska2009} for DLA systems at redshift 4.0--5.5. The simulations are compatible with the observational data, especially runs with no molecular cooling. \newline The main distinction between the models with and without molecular cooling is driven by the conversion of neutral Hydrogen to H$_2$, that becomes important in high density regions where new stars are being formed. Molecules boost the cooling of the surrounding gas and the formation of stars. As a result, the amount of atomic Hydrogen decreases. \newline

\begin{figure*}
	\includegraphics[scale=0.48]{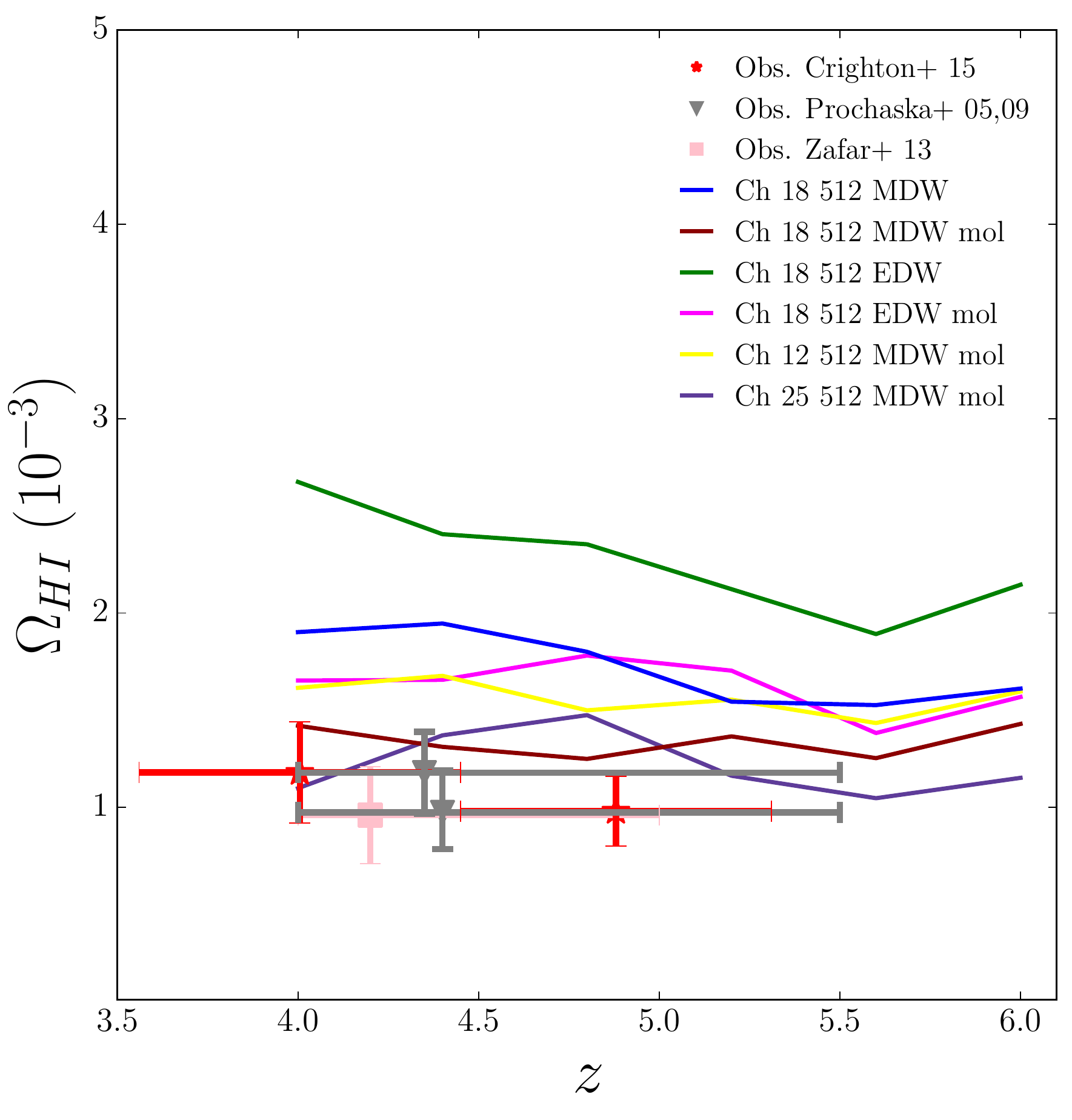}
	\includegraphics[scale=0.48]{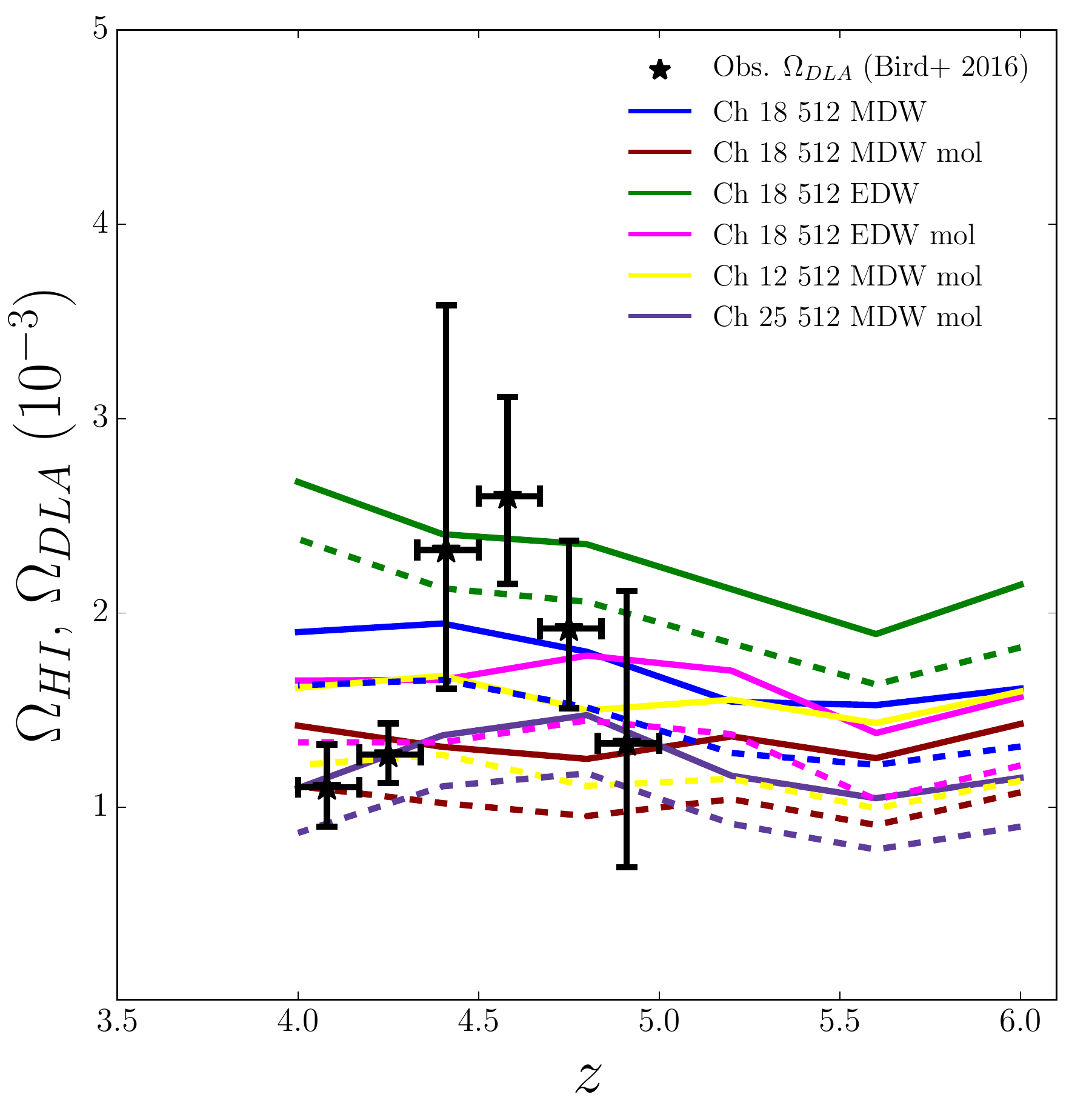}
        \caption{Cosmological mass density of \ion{H}{I}. In the left panel, we display the theoretical predition of $\Omega_{\ion{H}{I}}$ and compare with the data available at high redshift: \citet{prochaska2005, prochaska2009} in grey inverted triangles and \citet{zafar2013} in pink square and \citet{crighton2015} in red stars. The right panel shows $\Omega_{\ion{H}{I}}$ and $\Omega_{\text{DLA}}$ (solid and dashed lines, respectively). We compare the resulting $\Omega_{\text{DLA}}$ with recent data from SDSS \citep{bird2016} in black stars. In the observational work of \citet{crighton2015} it is proposed that more than 80\% of the contribution to the \ion{H}{I} cosmological mass density at $z<5$ comes from DLA systems. We confirm this assumption theoretically (Table~\ref{alphas}).}
    \label{fig:omega_hi}
\end{figure*}

At this point, we calculate the comoving mass density of \ion{H}{I}, using the following equation:
\begin{equation}\label{omegahi}
  \Omega_{\ion{H}{I}}(z)=\frac{H_0 m_{\ion{H}{I}}}{c \rho_{\text{crit}}}\int_{\text{N}_{\text{min}}}^{\text{N}_{\text{max}}} f_{\ion{H}{I}}(\text{N},z)\text{N} d\text{N},
\end{equation}
with $m_{\ion{H}{I}}$ the mass of the Hydrogen atom and $f_{\ion{H}{I}}(\text{N},z)$ defined in equation~\eqref{HI_CDDF}. In theory, the integral should be computed with an upper limit $\text{N}_{\text{max}} = \infty$, but in reality this value is set by the maximum column density detected. Thus, the limits considered for $\Omega_{\ion{H}{I}}$ are 12 $<$ $\log$ N$_{\ion{H}{I}}$ (cm$^{-2}$) $<$ 22. \newline We compare the pre\-dic\-tions from the theoretical models to observations at high redshift by \citet{prochaska2005}, \citet{prochaska2009}, \citet{zafar2013} and \citet{crighton2015} in the left panel of Figure~\ref{fig:omega_hi}. 

\noindent The amount of \ion{H}{I} at $z =$ 4 in the Ch 18 512 MDW mol (red line) and Ch 25 512 MDW mol (purple line) runs is in excellent agreement with the observational data, probably because the inclusion of molecular cooling prevents the production of too much neutral Hydrogen (by converting it into molecular Hydrogen). On the other hand, simulations without molecular cooling overproduce $\Omega_{\ion{H}{I}}$, especially in the EDW case (green line). All the models predict an increasing amount of \ion{H}{I} at $z>$ 5.5. However, we stress again that our assumption of a uniform UVB is no longer valid to describe the diffuse \ion{H}{I} in the IGM when approaching $z=$ 6. At these redshifts, our simulated $\Omega_{\ion{H}{I}}$ only describes the amount of neutral Hydrogen in collapsed/self-shielded systems. Please note that for this calculation we also extended the \ion{H}{I} self-shielding prescription of \citet{rahmati2013a}, which was originally proposed only up to redshift $z= $ 5. \newline

In order to estimate the contribution of DLAs to the overall \ion{H}{I} budget, the right panel of Figure~\ref{fig:omega_hi} compares the prediction of $\Omega_{\ion{H}{I}}$ for all the simulations (solid lines) with the DLA comoving mass density $\Omega_{\text{DLA}}$ (dashed lines), obtained by integrating the \ion{H}{I}--CDDF in the range 20.3 $<$ $\log$ N$_{\ion{H}{I}}$(cm$^{-2}$) $<$ 22. The theoretical estimations are compared with recent data from SDSS \citep{bird2016} in black stars. In this case, simulations without molecular cooling included are in better agreement with observations (as for the \ion{H}{I}--CDDF shown in Figure~\ref{fig:HICDDF2}).\newline We investigate the assumption $\Omega_{\ion{H}{I}} = \alpha\times\Omega_{\text{DLA}}$, where $\alpha$ is a factor that accounts for the numbers of systems that contain neutral Hydrogen with N$_{\ion{H}{I}} \geq$ 10$^{20.3}$ cm$^{-2}$. The values for $\alpha$ in each simulation (averaged in the redshift range 4--6) are in Table~\ref{alphas}. In the observational work of \citet{crighton2015}, the authors estimate a 20 per cent contribution from systems with column density below the DLA threshold (or 80\% contribution from DLAs) at $z<5$. This assumption seems to be validated with the numerical results here reported. \newline

\begin{table}
\caption{Contribution from DLA systems to $\Omega_{\ion{H}{I}}$.}
\label{alphas}
\centering
\begin{tabular}{lcc} 
\hline
Simulation &  $\alpha=\frac{\Omega_{\ion{H}{I}}}{\Omega_{\text{DLA}}}$ & \%\\ \hline
Ch 18 512 MDW  & 1.19 $\pm$ 0.03 &  80.9 \\ 
Ch 18 512 MDW mol & 1.25 $\pm$ 0.02 &  75.3\\ 
Ch 18 512 EDW  & 1.15 $\pm$ 0.03 & 84.9\\ 
Ch 18 512 EDW mol & 1.22 $\pm$ 0.02 & 77.8\\ 
Ch 12 512 MDW mol &  1.29 $\pm$ 0.04 & 70.4\\ 
Ch 25 512 MDW mol&  1.21 $\pm$ 0.02 &  78.7\\ 
\hline
\end{tabular}
\end{table}

\section{Discussion and summary}
\label{discussion}

The purpose of this work is to investigate the phy\-si\-cal environment of the IGM in proximity of galaxies at the end of the Epoch of Reionization by producing mock spectra and measuring ionic co\-lumn densities of metal transitions with a method that closely resembles current observational techniques. Taking advantage of a suite of simulations with a very robust model for chemical enrichment, SN--driven feedback that reproduces the cosmic SFR density and the galaxy stellar mass function at high redshift and a suitable co\-mo\-ving softening at the IGM scale (1--2 ckpc/$h$), we reproduce well some observables using metal and \ion{H}{I} absorption lines at high redshift. In particular, the comoving mass density $\Omega$ of \ion{C}{IV} and \ion{C}{II} in the redshift range of 4 $\le z \le$ 8, the \ion{H}{I} column density distribution function at $z =$ 4 and $\Omega_{\ion{H}{I}}$ \& $\Omega_{\text{DLA}}$ at $z<6$. In all cases, we compare the theoretical results with the latest release of observational data.

\subsection{Effect of wind feedback}

The simulations we used in this work are based on the analysis of \citet{tescari2014}, where there is an extensive discussion of the feedback mechanisms implemented and how they successfully reproduce the observations of the SFR at high redshift \citep[e.g.][]{Madau:2014bja}. When these models are used to study the evolution of different ionic species, we find that it is hard to disentangle the effect of different SN-driven galactic wind prescriptions for some of the statistics considered, in particular the \ion{C}{IV} and \ion{H}{I} column density distribution functions. The reasons are mostly numerical: probably higher resolution is needed to better resolve patches of enriched gas at relatively low-density in the CGM/IGM (as suggested by our resolution tests). Moreover, the SPH algorithm is not very effective in the mixing of metals. Therefore, only extreme feedback models would produce significant differences in the observables at high redshift. As a result, when calculating the column densities of metal absorption lines with Voigt profile fitting at $z >$ 5, it is not possible to distinguish between momentum--driven and energy--driven winds (MDW and EDW). The dispersion in the relationships among the metal absorption lines (see Figure~\ref{fig:correlations}) is too high to provide definitive hints on the evolution of the ionization states due to one particular wind model. \newline Nonetheless, a clear difference between the numerical estimation of the total Carbon content, $\Omega_{\text{C}}$, in EDW and MDW models arises at $z \leqslant$ 6 (Figure~\ref{fig:omega_c}). This is due to the fact that EDW quenches the overall star formation in galaxies more effectively than MDW, and consequently the production of Carbon in the simulation is 0.2--0.3 dex lower at $z =$ 4.

\subsection{Influence of the low temperature metal and molecular cooling}

As expected, the implementation of low temperature metal and molecular cooling affects mostly neutral Hydrogen statistics. In particular, molecular cooling suppresses \ion{H}{I} in regions where the majority of it is self-shielded, mainly because of the conversion of \ion{H}{I} to H$_2$ at high densities. This effect is important when computing the \ion{H}{I} column density distribution function and cosmological mass density. Interestingly, the introduction of molecular cooling improves the estimation of $\Omega_{\ion{H}{I}}$ at $z \sim$ 4--5 (left panel of Figure~\ref{fig:omega_hi}), whereas, when the \ion{H}{I}--CDDF is computed in the DLA regime, 20.3 $<$ $\log$ N$_{\ion{H}{I}}$ (cm$^{-2}$) $<$ 22, runs without molecular cooling provide a better match to new SDSS observations \citep{bird2016} of $\Omega_{\text{DLA}}$ (right panel of Figure~\ref{fig:omega_hi}) and the distribution function at $z =$ 4 (Figure~\ref{fig:HICDDF2}).\newline Once the feeback prescription is fixed, simulations with low-T metal and molecular cooling included produce slightly more \ion{C}{IV} than the other runs (Figure~\ref{fig:omega_c4} and left panel of Figure~\ref{fig:omega_c2}). The effect is less important for \ion{C}{II} and not visible for the total Carbon content and the dispersion in the relationships among the metal absorption lines (Figures~\ref{fig:omega_c} and \ref{fig:correlations}).

% At redshifts higher than $z = $ 4 the amount of neutral Hydrogen rises in all the simulations considered. At $z >$ 5, the model has not much predictive power, since it does not have a self-consistent calculation of the UVB and galaxies, the UVB is assumed, and the UVB is assumed to be uniform. In Figure~\ref{fig:omega_hi}, $\Omega_{\ion{H}{I}}$ is nearly flat at $z \sim$ 6, whereas if diffuse \ion{H}{I} is included, $\Omega_{\ion{H}{I}}$ should increase by a factor of 50. In addition, the self-shielded gas is not well described at high redshift, therefore Figure~\ref{fig:omega_hi} shows the density of neutral Hydrogen from the DLA systems contribution.

\subsection{Limitations}
\label{slimits}

Even though our models are in good agreement with observations of ionic species at high redshift, there are some caveats to consider. \newline First of all, radiative transfer effects are not included. The simulations are run in a relatively small volume (the maximum size con\-si\-de\-red is 25 cMpc/$h$), whereas a typical \ion{H}{II} bubble at the tail of EoR is larger than 100 cMpc/$h$. Therefore, we assume that our boxes represent a region of the Universe already reionized by a uniform HM12 UV ionizing background at $z\le8$. Although full RT calculations would be more accurate, this is a fair approximation to describe the ionization state of metals \citep{finlator2015}, since chemical enrichment at these redshifts occurs mostly inside and in close proximity of galaxies. On the other hand, even in small volumes the distribution of \ion{H}{I} optical depths cannot be properly described with a uniform UVB or a simple model that assumes galaxies and quasars as ionizing sources and uses a fixed mean free path for the ionizing photons \citep{becker2015}. For this reason, we have studied neutral Hydrogen statistics only at $z<6$. \newline A necessary step to complement the analysis of this work will be to vary the spectral hardness and the normalization of the UV background, in order to refine the calculation of the column densities of some ionic species, that are not well represented with the assumed uniform HM12 UVB. We plan to investigate this in a subsequent paper.\newline In recent years, numerical modeling of absorption features have improved enormously, but the current generation of simulations still struggles to reproduce low ionization states of metals at high redshift. This is mainly due to insufficient resolution, but it may also be influenced by the lack of a consistent implementation of self-shielding prescriptions for the ions. This work (like many others) just considers the effect of \ion{H}{I} self-shielding \citep{rahmati2013a}, but does not introduce any self-shielding of low ionization absorbers (which lay in clumpy structures), that has been proposed for DLA systems at lower redshift \citep{bird2015}.\newline Finally, there is a fundamental problem with the lack of reliable measurements of the mean normalized flux of the Lyman--$\alpha$ forest at $z \geq$ 5, when the QSOs spectra display very large Gunn--Peterson troughs, which leads to the absence of an effective parametrization of the optical depth of \ion{H}{I} to re-scale or calibrate the simulated Hydrogen and ion spectra. Different approaches have been used in the literature to reproduce the evolution of the metal ions independently from \ion{H}{I}. In this work, we calibrate our simulations using the \ion{C}{IV} column density distribution function at $z = $ 4.8 and 5.6.

\section{Conclusions}
\label{conclusions}

We have introduced high-resolution hydrodynamical simulations to study the physical and chemical state of the IGM at the end of the Epoch of Reionization. We have tested the effect of a low temperature metal and molecular cooling model and two prescriptions for galactic wind feedback on the evolution of metal absorption lines in the redshift range 4 $< z <$ 8. We have also studied the \ion{H}{I} cosmological mass density at $z<6$. Our theoretical predictions are consistent with the available observations at high redshift.

%A uniform HM12 UV ionizing background and Voigt profile fitting to calculate the column densities 
%of the ionization states of different metallic species are well motivated.\newline
%, despite of the limitations to resolve the absorption features at high redshift.\newline

\noindent  The evolution of the high and low ionization states of Carbon is the main focus of our work. The drop in the cosmological mass density of \ion{C}{IV}, $\Omega_{\ion{C}{IV}}$, from $z=$ 4 to 8 is due to the combined effect of a change in the ionization state of the gas and the decreased metallicity of the IGM (Figure~\ref{fig:omega_c4}). In fact, the total Carbon comoving mass density at $z=$ 8 is more than a factor of 10 lower than at $z=$ 4 (Figure~\ref{fig:omega_c}), and the simulations show a consistent transition to more neutral states of metals at high redshift (as shown in Figure~\ref{fig:omega_c2}). Most notably, the crossover between $\Omega_{\ion{C}{IV}}$ and $\Omega_{\ion{C}{II}}$ happens at $z\sim$ 6--6.5, in agreement with available observations and the current paradigm of the tail of Reionization. \newline In comparison to other numerical works in the literature, we are able to produce \ion{C}{IV} absorbers in the IGM with large column densities. This is mostly due to a combination of efficient galactic winds and a post-processing pipeline that mimics observational methods and, in particular, accounts for individual features to calculate the column densities from Voigt profile fits to the absorption lines. \newline The simulated \ion{C}{II} exhibits a bimodal distribution with large absorptions in and around galaxies, and some traces in the lower density IGM. These latter correspond to systems detected with current observations \citep{dodorico2013,becker2006}. We predict that, at high redshift, most of the high column density \ion{C}{II}, with $\log$ N$_{\ion{C}{II}}$ (cm$^{-2}$) $>$ 15, should lie in proximity of galaxies at impact parameters of order 20 ckpc/$h$ (Figure~\ref{fig:histo_c}) and has not been detected yet due to the low likelihood of reaching these overdense regions with lines of sight towards distant QSOs.\newline We have studied the column density relationships among different ionic species (Figure~\ref{fig:correlations}). High ionization states (like N$_{\ion{C}{IV}}$) are reasonably well described by our simulations with the adopted HM12 UV background. On the other hand, a comparison with observations in the literature shows that the low ionization states are not well represented, regardless of the feeback model implemented. \newline Finally, our simulations are in good agreement with observations of the \ion{H}{I} column density distribution function at $z=$ 4 (Figures~\ref{fig:HICDDF1} and \ref{fig:HICDDF2}) and the \ion{H}{I} cosmological mass density, $\Omega_{\ion{H}{I}}$, at 4 $<z<$ 6 (Figure~\ref{fig:omega_hi}). We validate the estimate made in the observational work of \citet{crighton2015} that DLA systems contribute to $\sim80$\% of $\Omega_{\ion{H}{I}}$ (see Table~\ref{alphas}). \newline

In a companion paper \citep{garcia2017a}, we have explored the likelihood of reproducing the observed Ly$\alpha$ emitter galaxy -- \ion{C}{IV} absorption pair detected by \citet{diaz2015} and studied the physical processes that produced the metal enrichment in the IGM at $z \geqslant$ 5.6. Future work will be focused on analyzing more metals and ions to have a broader perspective on the evolution of the IGM at the end of the EoR. Some of the column density relationships are extremely sensitive to the ionizing background, therefore varying the normalization and hardness of the UVB is the next step to obtain a more realistic description of this cosmic era.

\section*{Acknowledgements}

Parts of this research were conducted by the Australian Research Council Centre of Excellence for All-sky Astrophysics (CAASTRO), through project number CE110001020. The numerical simulations were run using the Raijin distributed-memory cluster from the National Computational Infrastructure (NCI) facility, post-processed with the Edward HPC machine from the University of Melbourne and analyzed with the gSTAR supercomputer at Swinburne University of Technology. This work was supported by the Flagship Allocation Scheme of the NCI National Facility at the ANU. The authors acknowledge CAASTRO for funding and allocating time for the project \textit{Diagnosing Hydrogen Reionization with metal absorption line ratios} (fy6) during 2015 and 2016.  L.A. Garc\'ia thanks Valentina D'Odorico and Neil Crighton for providing their observational data in private communications and Jeff Cooke for the insightful discussions. E. Ryan--Weber acknowledges ARC DP 1095600. We also thank the anonimous referee for providing valuable comments that considerably improved our manuscript and Klaus Dolag for his help with the technical details of the SPH code.

\bibliographystyle{mnras}
\bibliography{thebib}

\label{lastpage}

\end{document}